\documentclass[prd,twocolumn,showpacs,superscriptaddress,nofootinbib]{revtex4-1}
\usepackage{amsfonts}
\usepackage{tikz}
\usepackage{pgfplots}
\usepackage{amsmath}
\usepackage{amssymb}
\usepackage{bm}
\usepackage{dcolumn}
\usepgfplotslibrary{groupplots}
\usepackage{graphicx}
\usepackage{graphics}
\usepackage{latexsym}
\usepackage{rotating}
\usepackage[colorlinks=true]{hyperref}
\usepackage[all]{hypcap} 
\usepackage{xspace} 
\usepackage{xcolor}
\usepackage{mathrsfs}
\usepackage{siunitx}
\usepackage{multirow}
\usepackage{soul}
\usepackage{ulem}
\usepgfplotslibrary{fillbetween}
\definecolor{darkgreen}{rgb}{0.2,0.7,0.2}

\newcommand\be{\begin{equation}}
\newcommand\ee{\end{equation}}
\newcommand\bw{\begin{widetext}}
\newcommand\ew{\end{widetext}}

\newcommand{\bea}{\begin{eqnarray}}
\newcommand{\eea}{\end{eqnarray}}

\newcommand{\cm}{{\rm cm}}

\newcommand{\g}{{\rm g}}

\usepackage{cellspace} %
\newlength\figureheight
\newlength\figurewidth

\newlength\figureheightmed
\newlength\figurewidthmed

\newlength\figureheightlarge
\newlength\figurewidthlarge

\begin{document}
\title{Quantifying uncertainties in general relativistic magnetohydrodynamic codes}

\author{Pedro L. Espino}
\affiliation{Department of Physics, The Pennsylvania State University, University Park, PA 
16802, USA}
\affiliation{Department of Physics, University of California, Berkeley, CA 94720, USA}
\author{Gabriele Bozzola}
\affiliation{Department of Astronomy, University of Arizona, Tucson, 
  AZ 85721,USA}
\author{Vasileios Paschalidis}
\affiliation{Department of Astronomy, University of Arizona, Tucson, 
  AZ 85721,USA}
  \affiliation{Department of Physics, University of Arizona, Tucson, AZ 
85721, USA}

\begin{abstract}
  In this paper, we show that similar open-source codes for general
  relativistic magnetohydrodynamic (GRMHD) produce different results
  for key features of binary neutron star mergers. First, we present a
  new open-source version of the publicly available {\tt
    IllinoisGRMHD} code that provides support for realistic, finite
  temperature equations of state. After stringent tests of
  our upgraded code, we perform a code comparison between
  \texttt{GRHydro}, {\tt IllinoisGRMHD}, \texttt{Spritz}, and
  \texttt{WhiskyTHC}, which implement the same physics, but slightly
  different computational methods. The benefit of the comparison is
  that all codes are embedded in the \texttt{EinsteinToolkit} suite, hence
  their only difference is algorithmic.  We find similar convergence
  properties, fluid dynamics, and gravitational waves, but different
  merger times, remnant lifetimes, and gravitational wave phases.
  Such differences must be resolved before the post-merger dynamics
  modeled with such simulations can be reliably used to infer the
  properties of nuclear matter especially in the era of precision
  gravitational wave astronomy.
\end{abstract}

\date{\today} \maketitle

\section{Introduction}

The detection of gravitational waves (GWs) from the merger of a binary
neutron star (BNS)~\cite{TheLIGOScientific:2017qsa} has established
the need for a clearer understanding of BNS systems.  To gain insight
into the physics of BNS mergers, observations associated with such
events -- be they in the GW or electromagnetic (EM) spectrum -- must
be analyzed with either simplified analytical models which
parameterize the dominant physical phenomena in the system,
e.g.,~\cite{Raaijmakers:2021slr,
  Easter:2020ifj,Barbieri:2020ebt,Breschi:2021wzr,Nicholl:2021rcr} or
with the use of accurate numerical relativity (NR) simulations.
Parametric/phenomenological models of BNS mergers cannot reliably
capture the physics of the most extreme stages of the merger,
including the merger itself and the environment immediately following,
where the matter is in a highly dynamical state and the spacetime
curvature is the strongest. For a reliable, first-principles
understanding of these stages of BNS mergers, the use of NR is the
only recourse. Numerical simulations allow for the systematic
isolation of different physical phenomena/physics, which provides a
powerful tool for deducing the role of magnetic fields,
e.g.,~\cite{prs15,Ruiz:2017due,Ruiz_2016,Kiuchi:2017zzg,
  Shibata:2021bbj,Ciolfi:2020wfx,Mosta:2020hlh, Ciolfi:2020hgg},
equation of state (EOS) effects, e.g.,~\cite{Abdikamalov:2008df,
  Bauswein:2013jpa, Palenzuela_2015, Sekiguchi:2015dma,
  Lehner:2016lxy, Radice:2018pdn, Most:2018eaw, Dietrich:2018phi,
  Vincent:2019kor,Raithel:2021hye}, and neutrino transport, e.g.,~\cite{Sekiguchi:2015dma,Radice:2018pdn, Nedora:2019jhl,Foucart_2020}.

There are many important physical effects to consider in simulations
of BNS mergers (see~\cite{Paschalidis:2016agf, Baiotti2017, Duez2019,
  Radice_2020, Bernuzzi:2020tgt, Ciolfi:2020cpf} for reviews); such as
finite temperature EOSs and magnetic fields. Regarding the
use of realistic EOSs in BNS merger simulations, an important aspect
is the consistent evolution of the electron fraction, as it provides
crucial information for the description of nucleosynthesis processes
that take part during and after merger, and is necessary to fully
describe kilonova (KN) afterglows~\cite{Kasen_2013,Barnes:2013wka}.
Additionally, the thermodynamic properties of the post-merger remnant
and ejecta are important and not well approximated by cold EOSs (see,
e.g.,~\cite{Rezzolla:2017aly, raithel2021realistic}).  Magnetic fields
are also a crucial ingredient to consider in BNS merger
simulations. In particular, an area which remains poorly understood is
the effect of magnetic fields on ejecta properties.  Large scale
magnetic fields are important in the evolution of the post-merger
remnant and in the kilonova signal associated with BNS mergers, as
they are expected to power relativistic
jets~\cite{Blandford:1977ds,VanPutten:1999vda,Bucciantini:2011kx,
  Ruiz:2016rai,Mosta:2020hlh} and drive relativistic
outflow~\cite{Siegel:2014ita,Fernandez:2018kax,Metzger:2018qfl}.  The
interplay between finite temperature EOSs and strong magnetic fields
remains insufficiently characterized in BNS merger simulations. For
example, realistic EOSs have been shown to produce lower velocity
ejecta when compared to simple analytic
EOSs~\cite{Wanajo:2014wha,Sekiguchi:2015dma,Perego:2017wtu,
  Vincent:2019kor, Bernuzzi:2020tgt} while, on the other hand,
simulations which account for magnetic effects suggest that, if
magnetic fields are strong enough, shocked dynamical ejecta during
merger could be boosted to higher
velocities~\cite{Kiuchi:2017zzg,Radice:2018ghv,Shibata:2021bbj,
  Ciolfi:2020wfx,Mosta:2020hlh,Ciolfi:2020hgg}.  Most modern
simulations of BNS mergers consider only some of the aforementioned
physical effects, and there have only been a handful of simulations
which consider all effects at
once~\cite{Kiuchi:2017zzg,Shibata:2021bbj,Ciolfi:2020wfx,
  Mosta:2020hlh,Ciolfi:2020hgg,Sun:2022vri}. Even in the case of more complete simulations, all treatments of neutrino
transport to-date must make some level of approximation since solving
the full 6+1 dimensional neutrino transport equation at is not
possible with current schemes and computational resources.

 
There are currently available several codes with support to different levels
of physics. For
instance, many state-of-the-art codes with microphysical EOS
compatibility do not implement a constrained-transport treatment of
magnetic fields. While variety in the numerical methods and codes used in the study of
BNS mergers is a strong-point
from the perspective of understanding numerical systematic errors,
key differences between codes can lead to a range of predictions for the
relevant observables, depending on the physics included in simulations.
For example, although there is general agreement in the ejecta properties
predicted by different
codes~\cite{Palenzuela_2015,Sekiguchi:2015dma,Lehner:2016lxy,
Shibata:2017xdx,Perego:2017wtu,Bovard:2017mvn,Radice2018,
Nedora:2019jhl,Vincent:2019kor},
dynamical ejecta masses range over
$\SI{e-4}{M_\odot} \lesssim M_{\rm ej} \lesssim \SI{e-2}{M_\odot}$
while speeds range over $0.1c \lesssim v_{\rm ej} \lesssim 0.3c$,
depending on the total binary mass, EOS, and numerical methods
considered. Simulations with even the highest resolutions can have
numerical uncertainties of over
$40\%$~\cite{Foucart:2016rxm,Radice_2018,Shibata:2017xdx,Bernuzzi:2020txg}.
The need for accurate simulations and detailed estimates of the error budgets is
becoming more and more urgent as future GW detectors are expected to be more
sensitive than current systematic errors~\cite{Purrer:2019jcp}.
Future investigations of BNS mergers would ideally require a set of
catch-all, accurate, open source GRMHD codes which can reliably
simulate BNS systems while considering as many microphysical phenomena
as possible. At the very least, reliable future predictions using NR
codes require detailed comparisons, including cross-checks of the
results obtained using different codes and of the varied numerical
methods used within the codes themselves.

In order to begin addressing each of these needs, we first present a
new implementation of realistic, finite-temperature EOS support in
\texttt{IllinoisGRMHD}\footnote{Recently,~\cite{Werneck:2022exo} implemented a similar extension to \texttt{IllinoisGRMHD}.} (as extensively discussed below), and then
perform a systematic comparison between different open-source GRMHD
codes: \texttt{IllinoisGRMHD}~\cite{Etienne:2015cea}, \texttt{GRHydro}~\cite{Moesta:2013dna}, \texttt{Spritz}~\cite{2020zndo...3689751G}, and
\texttt{WhiskyTHC}~\cite{WhiskyTHC1, WhiskyTHC2, WhiskyTHC3}. These codes implement the same physics and have
largely overlapping numerical methods, e.g., they are all housed
within the \texttt{EinsteinToolkit}~\cite{Loffler2012ETK,
  zachariah_etienne_2021_4884780}, thereby we can isolate and compare
the impact on the evolution and uncover the existence of systematic
errors arising only from the GRMHD codes. This allows us to test the
robustness of certain results with respect to different computational
choices. In most cases, such choices are about failsafes for when
the main algorithms fail, as in the case of the
conservative-to-primitive scheme used in the artificial atmosphere. In our
comparison, we focused on BNS mergers and found that important
quantities disagree across the codes. For example, codes do not agree
on whether the remnant is stable or undergoes collapse very shortly after
merger, which has profound implications for EM observations. This
calls for more detailed studies and comparisons across different
codes.

This paper is split in two parts. First, we discuss in detail the
formalism employed by \texttt{IllinoisGRMHD}, the extensions required
to reach feature-parity with the other codes, and the tests we
performed. In particular, in Sec.~\ref{sec:CH7_basiceq} we review the
basic equations that are relevant to BNS systems, and provide detail
on their numerical treatment within \texttt{IllinoisGRMHD}. In
Sec.~\ref{sec:CH7_realisticEOS} we describe our methods for
implementing realistic, finite temperature EOS capability within
\texttt{IllinoisGRMHD}. In Sec.~\ref{sec:c2p_tests} we discuss
stringent dynamical tests of the extended version of {\tt
  IllinoisGRMHD} which cover a broad range of scenarios. Readers that
are familiar with GRMHD simulations may skip this part and focus on
the results. In the second part, starting from Section~\ref{subsec:G2BNS},
we discuss the results from our simulations of BNS mergers.

Our extension to \texttt{IllinoisGRMHD} to allow for the use of realistic,
finite-temperature EOSs is public. The extensions we have made to
\texttt{IllinoisGRMHD} are crucial for understanding the interplay between
magnetic fields and the EOS in the post-merger environment, and is a first step
to including neutrino transport schemes in the code. The current state of
\texttt{IllinoisGRMHD} (including the extensions we describe here) opens up many
avenues of investigation surrounding compact object mergers with strong magnetic
fields. For instance, the new code capabilities will make possible the
calculation of nucleosynthesis rates and help elucidate the role of
microphysical, finite temperature EOSs in BNS mergers with strong magnetic
fields, among many other interesting phenomena.

Throughout the work we use geometrized units, where $G=c=1$, unless otherwise
stated. In addition, in cases where we use logarithmic scales, we assume that
$\log\equiv\log_{10}$, unless otherwise noted. All visualizations and post-process
analyses in this work were carried out with the {\tt kuibit} software
package~\cite{kuibit}.

\section{Basic equations}\label{sec:CH7_basiceq}

Throughout this work we will predominantly work with numerical codes that
solve the Einstein field equations, 
\begin{equation}\label{eq:Einstein}
G_{\mu\nu} = 8\pi T_{\mu\nu},
\end{equation}
(where $G^{\mu\nu}$ and $T^{\mu\nu}$ are the Einstein and stress-energy 
tensors, respectively), coupled to the equations of ideal relativistic 
(magneto)hydrodynamics. In particular, we focus on 
the use of \texttt{IllinoisGRMHD} to solve the equations of ideal 
relativistic magnetohydrodynamics.
\texttt{IllinoisGRMHD} takes advantage of the 
Baumgarte-Shapiro-Shibata-Nakamura (BSSN) 
formulation~\cite{BSSN1,BSSN2,BSSN3} of the 3+1 
Arnowitt-Deser-Misner (ADM) formalism, which recasts the Einstein 
equations in the form of an initial-value problem (for more details, see textbooks
on the subject, e.g.~\cite{Alcubierre:2008it,2005LNP...673.....B,Baumgarte:2010nu,Shibata2016b}), in which the spacetime line element is given by
\begin{equation}
ds^2 = -\alpha^2 dt^2 + \gamma_{ij}(dx^i + \beta^i dt)(dx^j + \beta^j dt),
\end{equation}
where $\alpha$ is the lapse, $\beta^i$ is the shift vector, 
$\gamma_{\mu\nu} = g_{\mu\nu} + n_\mu n_\nu$ represents the 
induced metric on spacelike hypersurfaces, and 
$n^\mu=(1/\alpha, -\beta^i/\alpha)$ is the future-pointing unit vector 
orthogonal to each spatial slice.
The solution to Eq.~\eqref{eq:Einstein} involves the evolution of the 
magnetohydrodynamic variables that appear on the right hand sides of the 
ADM equations. An approach
which is well suited to the 3+1 formalism is
the Eulerian (or Valencia) formulation of relativistic 
hydrodynamics~\cite{Font:2008fka, 2013rehy.book}.
In this formulation, the evolution equations for the relevant fluid 
variables arise from several conservation 
laws, including the continuity equation
\begin{equation}\label{eq:rest_mass_cons}
\nabla_\mu (\rho_{\rm b} u^\mu) = 0,
\end{equation}
where $\rho_{\rm b}$ is the rest mass density and $u^\mu$ is the fluid 
four-velocity,
the lepton number conservation (when neutrino effects are ignored)
\begin{equation}\label{eq:ne_cons0}
u^\mu\nabla_\mu(Y_{\rm e}) = 0,
\end{equation}
which can be rewritten by use of Eq.~\eqref{eq:rest_mass_cons} as
\begin{equation}\label{eq:ne_cons}
\nabla_\mu(\rho_b Y_{\rm e} u^\mu ) = 0,
\end{equation}
where $Y_{\rm e}\equiv n_{\rm e}/n_{\rm b}$ is the electron fraction and 
$n_{\rm b}$ ($n_{\rm e}$) is the baryon
(electron) number density, the conservation of stress-energy
\begin{equation}\label{eq:Tmunu_cons}
\nabla_\mu T^{\mu\nu} = 0,
\end{equation}
and the homogeneous Maxwell's equations
\begin{equation}\label{eq:Maxwell_cons}
\nabla_\nu F^{*\mu\nu} = 0,
\end{equation}
where 
$F^{*\mu\nu} = \frac{1}{2} \varepsilon^{\mu\nu\alpha\beta}F_{\alpha\beta}$ 
is the dual to the electromagnetic field strength tensor 
$F_{\alpha\beta}$ and $\varepsilon^{\mu\nu\alpha\beta}$ is the rank-4 
Levi-Civita symbol. The matter variables are 
evolved using
Eqs.~\eqref{eq:rest_mass_cons}-\eqref{eq:Maxwell_cons} 
once they have been cast in flux-conservative form,
\begin{equation}\label{eq:evol_meta}
\partial_t \mathbf{C} + \mathbf{\nabla} \cdot \mathbf{F} = \mathbf{S},
\end{equation}
where $\mathbf{C}, \mathbf{F}$, and $\mathbf{S}$ are vectors which 
contain the conservative, flux, and source 
terms, respectively.
The vector of primitive variables $\mathbf{P}$
\begin{align}
    \mathbf{P} &= \begin{bmatrix}
           \rho_{\rm b} \\
           P \\
           v^i \\
           B^i\\
           Y_{\rm e}
         \end{bmatrix},
\end{align}
contains information on the physical state of the fluid, where $P$ is the 
fluid pressure, $v^i=u^i/u^0$ is the fluid three-velocity and $B^i$ 
are the spatial components of the magnetic field as measured by a normal 
observer.
 The conservative variables $\mathbf{C}$ are determined in terms of the 
 primitive variables, 
 the lapse $\alpha$, and the metric as
\begin{align}\label{eq:conservatives}
    \mathbf{C} &= \begin{bmatrix}
           \rho_* \\
           \tilde{\tau} \\
           \tilde{S}_i \\
           \tilde{B}^i\\
           \tilde{Y}_e
         \end{bmatrix} &= 
         \begin{bmatrix}
           \alpha\sqrt{\gamma}\rho_{\rm b}u^0 \\
           \alpha^2\sqrt{\gamma}T^{00} - \rho_* \\
           (\rho_* h + \alpha u^0 b^2)u_i - \alpha\sqrt{\gamma} b^0b_i\\
           \sqrt{\gamma}B^i\\
            \rho_*Y_{\rm e}
         \end{bmatrix},
  \end{align}
 where $b^\mu = B^\mu_{(u)} / \sqrt{4\pi}$ (where $B^\mu_{(u)}$ is the 
 magnetic field 
 as measured by an observer in the fluid rest-frame), $\gamma$ is the 
 determinant of the 3-metric, 
 and $h = 1 + \epsilon + P/\rho_0$ is the specific enthalpy (where 
 $\epsilon$ is the specific internal energy). 
 We note that \texttt{IllinoisGRMHD} uses the coordinate three-velocity, 
 $v^i \equiv u^i / u^0$, unlike many other evolution codes, which also 
 adopt the 
 Valencia formalism, but use the Eulerian three-velocity 
 $v^i_{(n)} = u^i/(\alpha u^0) + \beta^i / \alpha$~\cite{Banyuls:1997zz, 
 Anton:2005gi}, i.e., the velocity measured by normal observers.
 Additionally, 
 we note that \texttt{IllinoisGRMHD} works with
 an ideal fluid stress-energy tensor of the form
 \begin{equation}\label{eq:stress_energy}
 T_{\mu\nu} = (\rho_{\rm b} h + b^2) u_\mu u_\nu + (P + b^2/2)g_{\mu\nu} - b_{\mu}b_\nu.
 \end{equation}
 The evolution equations for the relevant fluid variables are determined 
 by using 
 the aforementioned conservation laws and casting them in the form of 
 Eq.~\eqref{eq:evol_meta}. 
 It is useful to pay special attention to the evolution of the magnetic 
 field, due to how it is
 treated in \texttt{IllinoisGRMHD}.
 Specifically, the evolution equation for the magnetic field is
 \begin{equation}\label{eq:B_evol}
 \partial_t \tilde{B}^i + \partial_j(v^j\tilde{B}^i - v^i\tilde{B}^j) = 0,
 \end{equation}
 where $\tilde{B}^i$ is the conservative variable corresponding to the magnetic field $B^i$, as defined in Eq.~\eqref{eq:conservatives}.
 The use of Eq.~\eqref{eq:B_evol} to evolve the magnetic field results in 
 terms that 
 violate the no-monopole constraint 
 ($\mathbf{\nabla} \cdot \mathbf{B} = 0)$, which is addressed  
 in \texttt{IllinoisGRMHD} by instead considering the evolution of the 
 vector potential $
 \mathcal{A}_\mu = \Phi n_\mu + A_\mu$ (where $A_i$ is the magnetic 
 vector potential and $\Phi$ is 
 electric scalar potential) and recovering the magnetic field as
 $\tilde{B}
 ^i = \varepsilon^{ijk} \partial_j A_k$. The evolution equation for $A_i$ is
 \begin{equation}\label{eq:A_evol}
 \partial_t A_i = \varepsilon_{ijk}v^j \tilde{B}^k - \partial_i(\alpha\Phi - \beta^{j}A_j).
 \end{equation}
\texttt{IllinoisGRMHD} works in the generalized Lorenz gauge
$\nabla_{\mu} \mathcal{A}^
\mu = \xi n_{\mu} \mathcal{A}^\mu$~\cite{Farris2012}, where $\xi$ is chosen such that the  
Courant-Friedrich-Lewy (CFL) condition corresponding to it is satisfied at all 
times given the grid choices. In the 
following we provide additional descriptions of the algorithms employed 
within \texttt{IllinoisGRMHD} for the implementation of finite 
temperature EOS compatibility. We direct the reader to~\cite{Etienne:2015cea} for a detailed description of all of the
additional algorithms employed within \texttt{IllinoisGRMHD}.


\section{Implementation of realistic equation of state compatibility within \texttt{IllinoisGRMHD}}\label{sec:CH7_realisticEOS}
The current open-source version of \texttt{IllinoisGRMHD} solves the equations of GRMHD by
assuming simple, analytic EOSs, such as $\Gamma$-law or piecewise polytropic EOSs with hybrid thermal effects. These EOSs can only provide a 
qualitative understanding of the state of matter during a BNS 
merger~\cite{Rezzolla:2017aly, Raithel:2021hye}. However, 
efforts to model parametrically both the thermal and the cold component
of the nuclear EOS in BNS mergers are under way, see, e.g.,~\cite{Raithel:2021hye,Most:2021ktk,Raithel:2022san,Raithel:2022nab}

The implementation of realistic EOSs within \texttt{IllinoisGRMHD} allows
us to understand, in a more detailed manner, the interplay between the EOS, 
thermal effects, and magnetic fields in these systems. Moreover, it is a crucial first step
toward implementing additional important microphysics within \texttt{IllinoisGRMHD}, such as neutrino
transport. The inclusion of realistic EOS capability within \texttt{IllinoisGRMHD} 
required two steps: the implementation of the evolution equation for the electron fraction and the
implementation of algorithms which can perform the non-trivial inversion of the evolved conservative
variables $\mathbf{C}$ to the physical primitive variables $\mathbf{P}$, which we refer to as 
conservative-to-primitive inversion. In the following
we discuss the implementation of the evolution equation for the electron fraction $Y_{\rm e}$ within our extended 
version of \texttt{IllinoisGRMHD}. 
We also discuss the implementation of state-of-the-art routines within 
\texttt{IllinoisGRMHD} for conservative-to-primitive inversion and present
relevant tests. 
For the remainder of this work, we refer to the 
currently available version of \texttt{IllinoisGRMHD} as \texttt{OriginalIllinoisGRMHD} (abbreviated 
as \texttt{OIL}). Our extended version of \texttt{IllinoisGRMHD} will be referred to 
as \texttt{MicrophysicalIllinoisGRMHD} (abbreviated as \texttt{MIL}). In cases where we discuss features which 
are common between the two codes, we will refer to them jointly as the \texttt{IllinoisGRMHD} code. 
Algorithmically, \texttt{OIL} and \texttt{MIL} are identical, 
except for the changes highlighted in this work which are required for 
realistic EOS compatibility.

\subsection{Evolution of the electron fraction}\label{subsec:CH7_Ye}
In flux-conservative form, the electron fraction evolution equation is 
\begin{equation}\label{eq:Ye_evol}
\partial_t(\tilde{Y}_{\rm e}) + \partial_j(v^j \tilde{Y}_{\rm e}) = 0.
\end{equation}
With the inclusion of Eq.~\eqref{eq:Ye_evol}, the full set of equations solved within 
\texttt{MIL} is
\begin{align}\label{eq:cons_evol}
    \partial_t \begin{bmatrix}
           \rho_* \\
           \tilde{Y}_{\rm e}\\
           \tilde{\tau} \\
           \tilde{S}_i\\
           A_i
         \end{bmatrix} &+ 
         \partial_j \begin{bmatrix}
				\rho_*v^j \\
				\tilde{Y}_{\rm e} v^j\\
           \alpha^2\sqrt{\gamma}T^{0j} - \rho_*v^j \\
           \alpha\sqrt{\gamma}T^j_i\\
           \alpha\Phi - \beta^{j}A_j
         \end{bmatrix} &=
         \begin{bmatrix}
				0 \\
				0\\
           s \\
           \frac{1}{2}\alpha\sqrt{\gamma}T^{\alpha\beta}g_{\alpha\beta,i}\\
           \epsilon_{ijk}v^j \tilde{B}^{k}
         \end{bmatrix},
  \end{align}
 where 
 \begin{equation}
 \begin{aligned}
   s = \alpha \sqrt{\gamma}[ (T^{00}\beta^i\beta^j + 2T^{0i} \beta^j +  T^{ij})K_{ij}\\ -   (T^{00}\beta^i +T^{0i}) \partial_i \alpha ],
   \end{aligned}
\end{equation}
 and $K_{ij}$ is the extrinsic curvature.
The evolution of $\tilde{Y}_{\rm e}$ follows that of the other conservative variables, which begins with 
the determination of initial conditions. Presently, we allow for $Y_{\rm e}$ 
to be initialized in two possible ways:
\begin{enumerate}
\item Linear profile: we set $Y_{\rm e} = \Upsilon\rho_{\rm b}$ (where $\Upsilon$ is a
constant that ensures proper dimensionality), such that the electron 
fraction profile is 
linear with respect to $\rho_{\rm b}$, in 
order to consider a simple profile where gradients are non-zero 
throughout 
the solution grid. This initialization 
is unphysical and only useful for testing the advection of $Y_{\rm e}$ 
in situations without 
realistic EOSs, i.e., passive advection of the
electron fraction.
\item $\beta$-equilibrium profile: we set $Y_{\rm e}(\rho_{\rm b})$ according to the conditions for
$\beta$-equilibrium in cold neutron star (NS) matter. 
This initial profile is suitable for realistic descriptions of isolated stars as well as 
for binaries that are initially separated at large
enough distances such that the components are cold. All of the BNS
initial data considered in our tests are built for quasi-equilibrium systems, in
which the assumption that the components are cold, $\beta$-equilibrated
stars is well justified. We use the same assumption of cold $\beta$-equilibrium to initialize
all other hydrodynamical variables at $t=0$. \\
\end{enumerate}
Once the initial data are specified for the primitive variables at all grid points, the conservative 
variables are obtained through the simple algebraic relations provided in Eq.~\eqref{eq:conservatives},
which provides $\mathbf{C}$ at all grid points. For the evolution of $\mathbf{C}$, we employ 3 ghost-zones  at the outer boundary of each
adaptive mesh refinement (AMR) grid. We fill all buffer zones at the refinement level boundary with data interpolated from neighboring rougher or
finer levels of refinement 
using standard prolongation or restriction operators, respectively. Filling in the buffer zones in this manner
results in $\mathbf{C}$ being either prolongated or restricted after 
calculation~\cite{Etienne:2015cea}.
To ensure consistency between $\mathbf{C}$ and $\mathbf{P}$, the 
primitives are recovered from the conservatives using a root-finding algorithm at $t=0$, which we discuss in 
Sec.~\ref{subsec:CH7_C2P}. Next, we 
evaluate the flux term $\mathbf{F}$ in preparation for the next time 
step. To this end, we must 
reconstruct the primitives between grid points (i.e, at cell interfaces). 
\texttt{IllinoisGRMHD} employs the piecewise parabolic method 
(PPM)~\cite{Colella:1982ee} for primitives 
reconstruction. Reconstruction is used to evaluate the primitives on the 
left and right interfaces of all grid points, $\mathbf{P}_{L,R}$,  in all 
directions. These interface values are then used to calculate the 
corresponding conservative variables 
at cell interfaces, which are in turn used to calculate the flux term in 
Eq.~\eqref{eq:Ye_evol} using a second-order, finite-volume, high-resolution shock-capturing (HRSC) scheme. The handling
of fluxes at grid interfaces $\mathbf{F}_{L,R}$ requires a solution to a 
Riemann problem. \texttt{IllinoisGRMHD} employs the standard Harten-Lax-vanLeer (HLL)~\cite{HLL}
approximate Riemann solver, where for a given direction the electron 
fraction flux is given by
\begin{equation}
F^{\rm HLL} (Y_{\rm e})= \dfrac{c^-F_R + c^+F_L - c^+c^-(\tilde{Y}_{{\rm e},R} - \tilde{Y}_{{\rm e},L})}{c^+ + 
c^-},
\end{equation}
where $c^{\pm} = \pm \text{max}\left(0, c^{R}_{\pm}, c^{L}_{\pm}\right)$ and $c^{L,R}_{\pm}$ are the maximum ($+$)
and minimum ($-$) characteristic speeds at the left (L) and right (R) cell interfaces
(see~\cite{ILGRMHD_2010,Etienne:2015cea} for further algorithmic details).
The derivatives of the fluxes are then determined and summed 
independently for each direction. For instance, the flux along the 
$x$-direction takes the form
\begin{equation}\label{eq:HLL_Ye}
{(\partial_x F^x)}_{ijk} = \dfrac{F^{HLL,x}_{i+\frac{1}{2}jk}(Y_{\rm e}) -
F^{HLL,x}_{i-\frac{1}{2}jk}(Y_{\rm e})}{\Delta x}.
\end{equation}
The fluxes along the $y$- and $z$-directions take a similar form, but we 
instead consider finite differencing along the $j$ and $k$ indices, 
respectively. 

The evolution equation for $Y_{\rm e}$ does not include source terms
in the absence of neutrinos, so the right-hand-side of
Eq.~\eqref{eq:HLL_Ye} is then passed to the method of lines
(\texttt{MoL}) thorn, which integrates the conservative variable
$\tilde{Y}_{\rm e}$ forward in time. At this point, the updated
conservative variables would be known at all grid points except the
outer boundary. The next step is to recover the primitive variables
given these evolved conservative variables (see
Sec.~\ref{subsec:CH7_C2P}).  After the primitives have been recovered,
they are checked for physicality and marginally modified if they are
outside of their physical ranges~\cite{Etienne:2011ea}. For example, in the case
of the electron fraction, we check that
\begin{equation}\label{eq:ye_physical}
Y_{\rm e, \text{lower}} \leq Y_{\rm e} \leq Y_{\rm e, upper},
\end{equation}
where $Y_{\rm e, \text{lower}}$ ($Y_{\rm e, \text{upper}}$) corresponds 
to the lowest (highest) value for 
$Y_{\rm e}$ available in an EOS table.
Next, outer boundary conditions are placed on the recovered primitives to 
fill the necessary 3 
ghost-zones in each direction. We apply zero-derivative outflow outer 
boundary conditions as 
described in~\cite{Etienne:2015cea}. Up to this point, the primitives
$\mathbf{P}$ are 
known at the new time step on all grid points. The final step we take is 
to recompute the conservatives on all grid points using 
Eq.~\eqref{eq:conservatives} for 
consistency between $\mathbf{P}$ and $\mathbf{C}$, and the evolution 
algorithm is allowed to proceed.

\subsection{Conservative-to-primitive solvers}\label{subsec:CH7_C2P}
\begin{figure*}[htb]
\centering
\includegraphics[width=16cm]{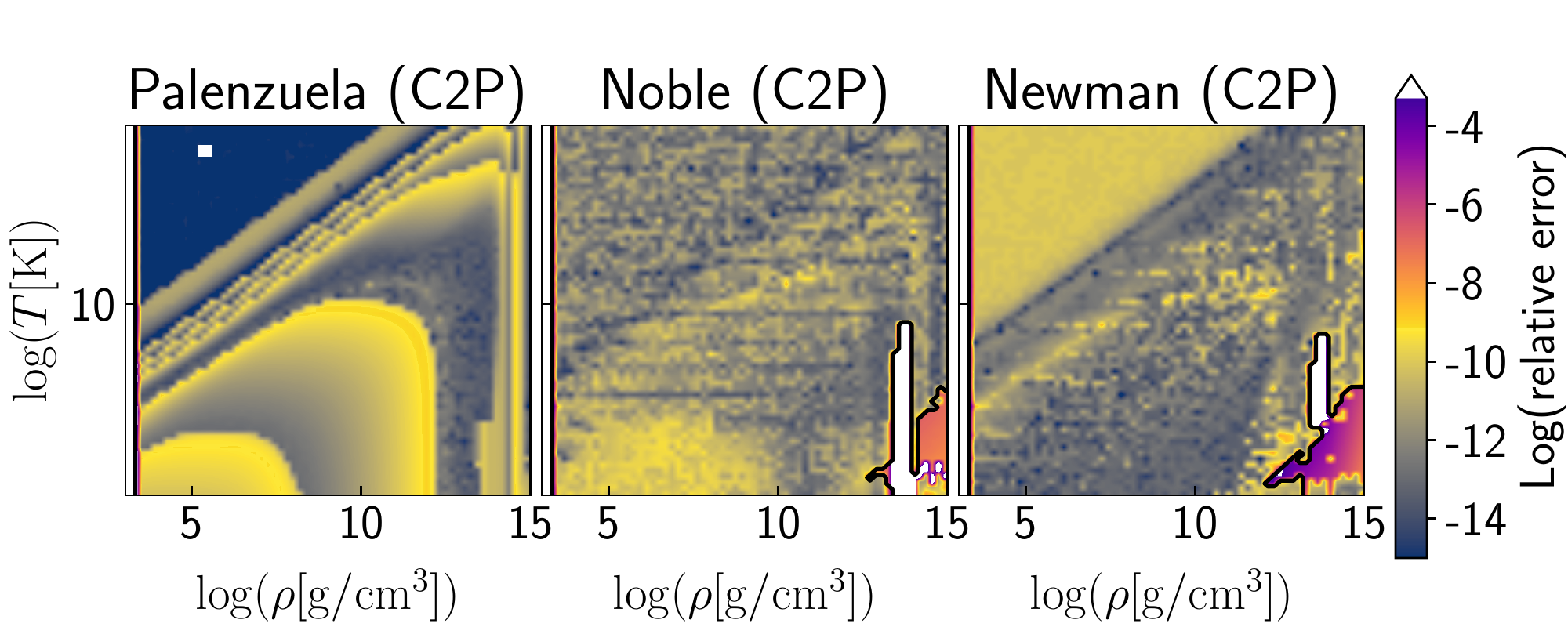}
\includegraphics[width=16cm]{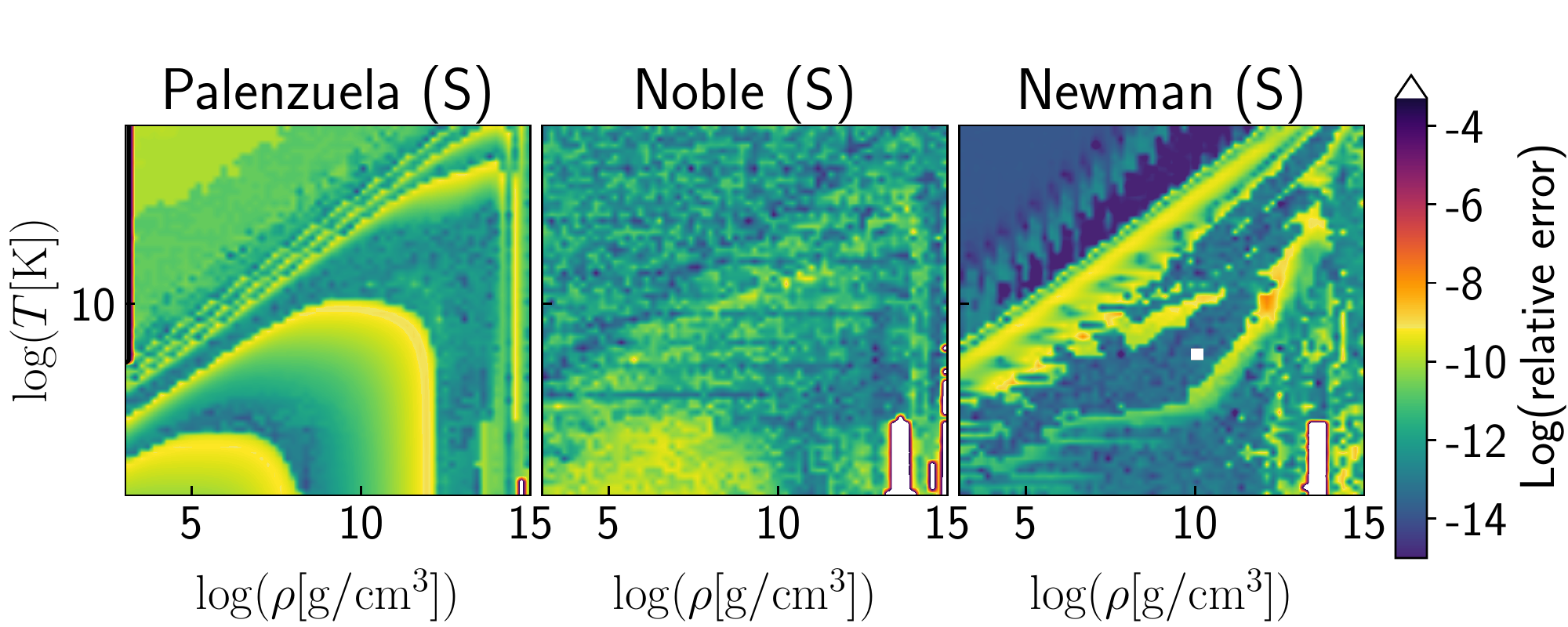}
\caption{\textit{Top panel:} Relative error in the recovery of primitives 
for a selection of the solvers 
originally implemented in~\cite{Siegel:2017sav} and ported to
the\texttt{EinsteinToolkit} within 
the \texttt{ConservativeToPrimitive} thorn (labeled C2P). 
We test the LS220 EOS~\cite{stellarcollapse, LS220}
with three of the available solvers. For all tests, we fix a subset of 
the primitives and
provide random initial data for the remainder. 
Specifically, we fix 
$P_{\rm mag}/P=10^{-3}$, $W=2$, and $Y_{\rm e} = 0.1$ while scanning the 
physically
allowed range for $T$ and $\rho_{\rm b}$. \textit{Bottom panel}: Same as 
the top panel, but using the original 
implementation of each solver in the code packaged 
with~\cite{Siegel:2017sav} (labeled S).}%
\label{fig:test_con2prim}
\end{figure*}

At every step of the evolution an inversion from the evolved conservative variables $\mathbf{C}$ 
to the physical primitive variables $\mathbf{P}$ 
is required to know the state of the fluid. 
Eq.~\eqref{eq:conservatives} presents a system of 
non-linear algebraic expressions which can be solved for 9 relevant fluid variables
($\rho_{\rm b}$, $v^i$, $B^i$, $Y_{\rm e}$, and either $h$, or $\epsilon$). These 9 variables, along with an EOS, in turn provide 
all of the information required to determine $\mathbf{P}$, along with other fluid 
variables of interest. For example, a solution to Eq.~\eqref{eq:conservatives}
can provide the 5 main variables ($\rho_0$, $v_i$, $\epsilon$) and, trivially, 
$B_i$ and $Y_{\rm e}$. We can then determine the remaining variable, $P$ with the use of 
an EOS and incidentally obtain information on other physical variables such as the temperature $T$ 
and specific entropy $s_{\rm b}$. 
As we require a solution for 5 \emph{main} variables in order to determine $\mathbf{P}$, the primitives inversion
problem is fundamentally a non-trivial 5D problem that cannot be solved analytically. 5D schemes which solve
Eq.~\eqref{eq:conservatives} were originally implemented in early MHD 
codes such as \texttt{HARM}~\cite{2003ApJ...589..444G}. However, these schemes were eventually
found to be inefficient and inaccurate, which led to the development of methods which
solve for only two auxiliary variables~\cite{Noble:2005gf} and thereby reduce the dimensionality
of the problem to 2D. The dimensionality of the problem can be further reduced to 1D;
modern 1D algorithms which provide reliable and efficient solutions 
have been developed~\cite{doi:10.1137/140956749, Palenzuela_2015} and are widely used 
in GRMHD codes.

In order to consider strongly magnetized systems which include realistic descriptions of the 
dense matter EOS, we implement state-of-the-art conservative-to-primitive solvers within
\texttt{MIL}. Our implementation includes porting the solvers discussed 
in~\cite{Siegel:2017sav} for use in the
\texttt{EinsteinToolkit}, packaged within a new thorn \texttt{ConservativeToPrimitive} 
which can interface with \texttt{MIL} and its associated 
thorns, but also works as a standalone thorn which can be used with other GRMHD codes that operate within the
\texttt{Cactus} infrastructure. 
We focus on a subset of the solvers implemented in~\cite{Siegel:2017sav}, due
to their reliability, speed, and algorithmic similarity to the original solvers used in
\texttt{OIL} (see~\cite{Kastaun:2020uxr} for another possible approach). In particular we focus on the 2D solver of~\cite{Noble:2005gf} (which
we label \texttt{Noble}), the 1D solver of~\cite{Palenzuela_2015} 
(which we label~\texttt{Palenzuela}), and the 1D solver of~\cite{doi:10.1137/140956749} 
(which we label \texttt{Newman}). We note that the \texttt{Noble} 
solver uses the same algorithm as the solvers in the original version 
of \texttt{IllinoisGRMHD} and that other GRMHD codes with realistic EOS capability rely on 
the \texttt{Palenzuela} algorithm~\cite{2020zndo...3689751G,Most:2019kfe, 
Kalinani:2021ofv}.
We refer the reader to~\cite{Siegel:2017sav} for a review of the algorithms used in each solver 
within \texttt{ConservativeToPrimitive}\footnotemark[1]. 

\footnotetext[1]{We point out that~\cite{Siegel:2017sav} appears to
have typographical errors in the algorithm description for the \texttt{Newman} solver, when compared to the
original paper of~\cite{doi:10.1137/140956749}. In particular, there is a difference in the calculation
of the auxiliary variable $\mathcal{M}^2$ (see Eq. (47) of~\cite{Siegel:2017sav} as compared to 
Eq.~(4.7) of~\cite{doi:10.1137/140956749}), where~\cite{doi:10.1137/140956749} correctly calculates it as
$\mathcal{M}^2 = m_{i}m^i = \tilde{S}_i\tilde{S}^i$,
where $\tilde{S}_i$ is the conservative variable associated with the momentum density which appears in
the left-hand-side of Eq.~\eqref{eq:cons_evol}. Eq. (47) of~\cite{Siegel:2017sav} 
inconsistently calculates this variable as $\mathcal{M}^2 = {(B^{i}v_i)}^2/\sqrt{\gamma}$, where $\gamma$ is
the determinant of the 3-metric $\gamma_{ij}$, $B^i$ is the magnetic field, and $v_i$ is the 
fluid 3-velocity. We also note that the first term 
of Eq. (55) in~\cite{Siegel:2017sav} misses a factor of the 
auxiliary variable $a$, when compared to the analogous Eq.~(5.11) in~\cite{doi:10.1137/140956749}.
We point out that despite these typographical errors, the numerical implementation of the \texttt{Newman}
solver within the code of~\cite{Siegel:2017sav} is consistent with the correct algorithmic steps
presented in~\cite{doi:10.1137/140956749}.}

We employ a set of preliminary tests to confirm that our port of these solvers
to the \texttt{EinsteinToolkit} behaves as intended. In particular, we test the aforementioned 
solvers using the same tests as~\cite{Siegel:2017sav} where a set of primitives $\mathbf{P}$
are initialized, randomly perturbed, used to calculate a set of 
conservatives $\mathbf{C}$,
and finally recovered into a new set $\mathbf{P}$'. 
The primitives recovery for each solver is then assessed by considering the relative error between the original 
set $\mathbf{P}$ and the recovered set $\mathbf{P}'$ (along 
with other diagnostics including the number of interpolation calls to the EOS table and number of algorithm 
iterations). 
In these tests, a subset of the primitives is varied over the physically allowed range while holding
others constant. We focus on the case of the LS220 realistic, finite temperature, tabulated 
EOS~\cite{stellarcollapse, LS220} and employ tests where we prescribe the ratio of 
magnetic to fluid pressure $P_{\rm mag}/P = b^2/(2 P)=0.001$, the Lorentz factor $W=2$, and the electron fraction
$Y_{\rm e}=0.1$, while scanning over the allowed range of rest mass density $\rho_{\rm b}$ and temperature $T$ for 
this EOS\@. For each value of $\rho_{\rm b}$ and $T$ that we consider, we determine the relative error between the
original and recovered set of primitives.
We present the results for this set of tests
in the top panel of Fig.~\ref{fig:test_con2prim}. As a comparison, we also show the 
results of these tests using the implementation of the solvers within the code of~\cite{Siegel:2017sav} in the 
bottom panel of Fig.~\ref{fig:test_con2prim}.

For all tests, we find very good agreement between the original solvers and our implementation of them
within the $\texttt{EinsteinToolkit}$. Points on Fig.~\ref{fig:test_con2prim} which 
appear in white indicate the failure of a given solver to recover a consistent set of
primitives. Points at which the solvers fail which are surrounded by successful recovery 
are usually due to the random perturbations we induce on $\mathbf{P}$ 
before recovery, which occurs for both the original code 
of~\cite{Siegel:2017sav} and our implementation within 
{\tt ConservativeToPrimitive} (see 
the top left and lower right panels of Fig.~\ref{fig:test_con2prim}).
However, there are regions of variable space where solvers consistently fail regardless 
of the initial perturbations (in both our implementation of the solvers within 
\texttt{ConservativeToPrimitive} and the original implementation in~\cite{Siegel:2017sav}). 
The most reliable solver across variable space is the \texttt{Newman} solver
although, as indicated by the rightmost panels of Fig.~\ref{fig:test_con2prim},
it may fail for cold matter ($T \lesssim \SI{10}{MeV}$) in the high density 
regime $\rho_{\rm b} \approx \SI{e14}{\g\per\cm\cubed}$, which is a highly relevant area
of variable space for BNS mergers. 
It is important to note that regions where a given solver may fail are often well covered by at least
one of the other solvers. Therefore, in practice the best use of these solvers 
should employ a hybrid approach, where a solver which fails in a given region of
variable space is substituted by a solver which provides reliable solutions
in that same region. In the following we discuss 
the choice of solvers we employ moving forward.

\subsection{Solution algorithm}
The preliminary tests presented in Fig.~\ref{fig:test_con2prim} suggest 
that some solvers are better
suited for systems involving neutron stars with realistic EOSs over others. 
Nevertheless, we note that the use of \emph{any} solver is still likely 
to produce robust solutions.
Therefore, the chief property that we use in determining the primary 
solver to utilize is the performance speed and agreement 
with the results of the \texttt{OIL} code.
For realistic EOSs, the bottleneck in limiting solver speed is typically the number of
times the solver must interpolate the EOS\@. Among the solvers
in \texttt{ConservativeToPrimitive},
the \texttt{Noble} solver requires the fewest such EOS calls, resulting 
in the most efficient solutions~\cite{Siegel:2017sav}. 
Nevertheless, the \texttt{Palenzuela} and \texttt{Newman} solvers require 
a similar number of EOS calls
to \texttt{Noble}, and the \texttt{Palenzuela} solver has the added 
benefit of strong agreement with
\texttt{OIL} and of reliable coverage across densities and temperatures,
as suggested by the leftmost panels of Fig.~\ref{fig:test_con2prim}. 
Due to the reliability and 
agreement with \texttt{OIL}, and because the performance is not much 
worse than other solvers, we consider
the \texttt{Palenzuela} solver as the optimal choice for our primary 
solver. In the remainder of this work, 
we adopt the following solution algorithm for conservative-to-primitive 
inversion in \texttt{MIL}
\begin{enumerate}
\item We employ the \texttt{Palenzuela} solver in our first attempt at a 
solution. 
We find that the \texttt{Palenzuela}
solver is robust for both analytic and microphysical EOSs and comparably 
efficient to other solvers.
\item If the \texttt{Palenzuela} solver fails, as it is disposed to for 
highly relativistic matter, we 
try to recover a solution with either the \texttt{Newman} or \texttt{Noble}
solvers.
The {\tt Noble} solver is robust for highly magnetized flows and highly 
efficient, but begins to fail at high Lorentz factors.
The \texttt{Newman} solver is reliable for both highly magnetized and 
highly relativistic flows. 
\item If all of the aforementioned solvers fail we impose a fix of the 
velocities, as done in the 
\texttt{OIL} code~\cite{Etienne:2015cea}, which assumes a 
cold EOS ($T=\SI{0.1}{MeV}$); we refer 
to this fix as the `Font' fix,
which has been generalized from its original form in \texttt{OIL} for 
generic EOSs.
\item Finally, if all other recovery attempts have failed, it is likely 
that the fluid
element is at low densities, so we set the recovered primitives to values 
which are sensible
for atmospheric conditions (cold temperature $T_{\rm atm}=\SI{0.1}{MeV}$, 
low density 
$\rho_{\rm b, atm} \leq 10^{-7}\rho_{\rm b, max}$, fixed electron 
fraction $Y_{\rm e, atm} = 0.25$,
and a pressure which is consistent with the EOS 
$P_{\rm atm}=P(\rho_{\rm b, atm}, T_{\rm atm}, Y_{\rm e, atm})$).
 We note 
that the other open-source codes we consider, 
with the exception of {\tt GRHydro} which uses the same treatment as 
{\tt MIL}, only fix 
atmospheric values 
of $\rho_{\rm b}$ and $T$, but not $Y_{\rm e}$. We 
emphasize, however, 
that the values of the primitive variables in the 
atmosphere are generally dynamically unimportant, and are 
set for numerical stability.
\end{enumerate}
We find that the above solution algorithm leads to a robust and efficient 
evolution across
the physically allowed range of variables for a finite temperature, 
realistic EOS\@. We note that
the order of solvers used is completely customizable at run-time and 
can be changed dynamically for any 
simulation, and that any ordering
of the \texttt{Palenzuela}, \texttt{Noble}, and \texttt{Newman} solvers 
results in a reliable solution.  
The order of solvers chosen here is optimal when considering efficiency 
and reliability for finite temperature, 
tabulated EOSs. In the remainder of this work, we utilize 
the \texttt{MIL} code with the above solution algorithm, unless otherwise 
noted.

\section{Dynamical tests of \texttt{MIL} and code comparisons}\label{sec:c2p_tests}

\subsection{Summary of tests, initial data, and grid hierarchies}\label{subsec:test_summary}
\begin{table*}[htb]\label{tab:grids_C2Ptests}
\centering
\caption{Grid hierarchy and simulation settings for the tests presented 
in Sec.~\ref{sec:c2p_tests}, in the case of the lowest-resolution grids 
employed. For each case we list the test 
name, the finest-level grid resolution $dx_{\rm fin}$, the number of 
refinement levels (RL), the number of grid points per NS radius (GP), and
the half-side length of each nested grid (HSL) listed from finest to 
coarsest. We also list the evolution codes (Evol.\ code),
Riemann solver (Riemann sol.), 
reconstruction scheme (Recon.\ scheme), and conservative-to-primitive
error tolerance (C-to-P tol.) used 
in each simulation. In terms of the Riemann solver, we use either the 
Harten-Lax-vanLeer (HLL)~\cite{HLL} or 
Harten-Lax-vanLeer-Einfeldt (HLLE)~\cite{doi:10.1137/0725021}, depending 
on the algorithm available in each code.  }
  \begin{tabular}{l | c c c c | c c c c }\hline  \hline
Test        & $dx_{\rm fin}$ & RL & GP & HSL ($M_\odot$) & Evol.\ code &
Riemann sol. & Recon.\ scheme & C-to-P tol. \\\hline
G2TOV       & 0.25   & 1  & 32 & $(10.0)$ & \texttt{MIL} & HLL & PPM & $10^{-8}$\\ 
& & & & & \texttt{GRHydro} & HLLE & PPM & $10^{-8}$ \\\hline
$\rm LSTOV$ & 0.111  & 3  & 64 & $(8.9,17.8,35.6)$ & \texttt{MIL} & HLL & PPM & $10^{-8}$ \\ 
 & &  & & & \texttt{Spritz} & HLLE & PPM & $10^{-8}$ \\ \hline
G2BNS       & 0.155  & 7  & 40 & $(13.0,17.9,26.1,60.0,120.0,240.0,396.0)$ & {\tt OIL} & HLL & PPM & $10^{-8}$\\ 
& & & & & \texttt{MIL} & HLL & PPM & $10^{-8}$\\ 
& & & & & \texttt{Spritz} & HLLE & PPM & $10^{-8}$\\ 
& & & & & \texttt{GRHydro} & HLLE & PPM & $10^{-8}$\\ 
& & & & & \texttt{WhiskyTHC} & HLLE & MP5 & $10^{-8}$\\
\hline
LSBNS       & 0.14  & 7  & 50 & $(8.7,17.4,34.8,69.6,139.2,278.4,556.8)$ & \texttt{MIL} & HLL& PPM & $10^{-8}$\\ 
& & & & & \texttt{GRHydro} & HLLE & PPM& $10^{-8}$\\ 
& & & & & \texttt{WhiskyTHC} & HLLE & MP5& $10^{-8}$\\ 
\hline
\end{tabular}\\
 \end{table*}
In this section we focus on code comparisons between {\tt MIL} and other 
open-source codes GR(M)HD codes. As tests of 
\texttt{MIL}, we employ a number of evolutions for which the system 
behavior is well understood. 
Where relevant, we compare the results of the \texttt{MIL} code to other 
publicly available codes with similar capabilities. In particular we 
compare with the \texttt{GRHydro}~\cite{Moesta:2013dna},
{\tt Spritz}~\cite{2020zndo...3689751G}, 
{\tt WhiskyTHC}~\cite{WhiskyTHC1, WhiskyTHC2, WhiskyTHC3}, and 
{\tt OIL}~\cite{Etienne:2015cea} codes.

We present the results of these tests as follows: 
\begin{enumerate}
\item G2TOV\@: We first discuss tests which employ a TOV star built with a
$\Gamma=2$ polytrope, 
which we label test $\text{G2TOV}$. For test $\text{G2TOV}$ we compare 
the solution to results obtained with the \texttt{GRHydro} code. This 
test was designed to mirror those considered in~\cite{Etienne:2015cea}.
\item LSTOV\@: Next, we discuss tests where we consider a magnetized TOV
star constructed using the finite temperature, tabulated LS220 
EOS~\cite{stellarcollapse, LS220}, which we label test $\text{LSTOV}$. 
For this test we compare our solutions to those obtained with the 
publicly available \texttt{Spritz} code, 
which uses an implementation of the \texttt{Palenzuela} solver for
finite temperature EOSs.
\item G2BNS\@: Next, we test our code in the case of a BNS merger, assuming
a $\Gamma=2$ polytropic EOS\@. We compare the solution against the results
of the {\tt GRHydro}, {\tt Spritz}, {\tt WhiskyTHC}, and {\tt OIL} 
codes. We 
consider both magnetized and unmagnetized systems, and refer 
to this test as G2BNS\@.
\item LSBNS\@: Finally, we consider a BNS built using the LS220 EOS\@.
Along with the {\tt MIL} code, we simulate this system with the 
{\tt GRHydro}, and {\tt WhiskyTHC} codes.
The binary configuration is built such that there is a transient hypermassive neutron star (HMNS)
post-merger remnant, which allows us to compare properties of the merger 
remnant between codes. We also consider evolution with the {\tt Spritz} code, but we do not include those results, because the public version of the code is still 
under development for cases with tabulated EOS support in the context of binary neutron star mergers.
We label this test as LSBNS\@.
\end{enumerate}

The use of \texttt{MIL} and \texttt{ConservativeToPrimitive} requires the 
initialization of three new 
required variables when evolving finite temperature, realistic EOSs. 
Namely, we must initialize the 
specific internal energy $\epsilon$, temperature $T$, and electron fraction 
$Y_{\rm e}$ such that they are consistent 
with the initial data. For tests which employ a cold polytropic EOS we 
initialize the specific internal energy
 as
\begin{equation}
\epsilon = \dfrac{P}{(\Gamma - 1)\rho_0}.
\end{equation}
For polytropic EOSs, the initialization of $T$ and $Y_{\rm e}$ are 
independent to that of $\epsilon$ and do not play a role in the fluid 
evolution, so we simply initialize 
$Y_{\rm e}$ based on the first type of profile listed in 
Sec.~\ref{subsec:CH7_Ye} and fix $T=\SI{0.1}{MeV}$, although the temperature is a passive variable in such tests.
For tests which employ realistic, finite temperature EOSs, we are
interested in equilibrium initial data, which requires cold, uniform 
temperatures and $\beta$-equilibrium. 
To construct cold, $\beta$-equilibrium initial data, we build barotropic 
tables that provide
$P(\rho_{\rm b})$.
Due to the fact that \texttt{ConservativeToPrimitive} makes use of EOS 
tables in the format of
the \texttt{StellarCollapse} repository~\cite{stellarcollapse} via the 
EOS driver thorn \texttt{EOS{\_}Omni}, 
we construct cold, $\beta$-equilibrium tables required for initial data 
using 
the \texttt{StellarCollapse} tables. 
The \texttt{StellarCollapse} tables provide several fluid variables 
(including the pressure $P$, 
specific internal energy $\epsilon$, and constituent chemical potentials 
$\mu_i$, 
where $i=p,n,e$ is an index over the particle species for protons, 
neutrons, and electrons, respectively) 
at discrete values of the rest mass density $\rho_{\rm b}$, electron 
fraction $Y_{\rm e}$, and temperature $T$. 
Each of the fluid variables may be interpolated to arbitrary values of 
the triplet $(\rho_{\rm b}, Y_{\rm e}, T)$. 
To extract the cold, $\beta$-equilibrated, barotropic 
functions 
required for initial data ($P(\rho_{\rm b})$, 
and $\epsilon(\rho_{\rm b})$), we take the following steps:
\begin{enumerate}
\item We fix the value of $T=T_{\rm cold}=\SI{0.1}{MeV}$. The lowest 
possible temperature values in these tables
are typically $T=0.01-\SI{0.05}{MeV}$, but entries near the table limits 
are not finely sampled,
which often leads to interpolation errors in the primitives recovery 
process. As such, we employ a finite 
temperature which is still cold compared to the Fermi energy 
but avoids the table boundaries.
\item For a given value of $\rho_{\rm b}=\rho_{\rm b, curr}$, we scan the 
available range of $Y_{\rm e}$ and 
evaluate the chemical potentials relevant for beta equilibrium 
($\mu_i(Y_{\rm e},\rho_{\rm b, curr},T_{\rm cold})$). 
\item We employ root-finding to 
locate the value of $Y_{\rm e}$ which corresponds to $\beta$-equilibrium, 
such that
\begin{equation}
\mu_n - \mu_p - \mu_e = 0.
\end{equation}
At this value of $Y_{\rm e}=Y_{\rm e, \beta}$, 
we record the pressure and specific internal energy.
We continue the algorithm from step 2 above, scanning the available range 
of $\rho_{\rm b}$ and 
thereby constructing tabulated functions for the barotropic pressure $P$ 
and specific internal energy $\epsilon$
as functions of the rest mass density $\rho_{\rm b}$ 
for $\beta$-equilibrium matter.
\end{enumerate}
 
In Tab.~\ref{tab:grids_C2Ptests} we present the grid hierarchy 
corresponding to each dynamical test
we consider. For each case we list the test name, the finest-level grid
resolution $dx_{\rm fin}$,
the number of refinement levels (RL), the number of
grid-points per NS radius (GP), the half-side length of each nested grid
(HSL) listed from finest to
coarsest, and the evolution codes used.
In all cases
we evolve the spacetime using the {\tt McLachlan} spacetime evolution code
within the {\tt EinsteinToolkit} (via the {\tt ML\_BSSN}
thorn)~\cite{Brown2009, Reisswig_2011GWs}, unless otherwise noted.
{\tt ML\_BSSN} solves the Einstein equations within the BSSN formulation of
the ADM 3+1 formalism.
Our gauge conditions consist of the ``1+log'' slicing condition for the
lapse~\cite{Bona:1994dr}
and the ``Gamma-driver'' condition for the shift~\cite{Alcubierre:2002kk}.
Integration in time is carried out using a fourth-order accurate Runge-Kutta (RK4)
scheme, using the {\tt MoL} thorn, with a Courant factor of 0.5, unless
otherwise stated.

\subsection{Diagnostics}
We employ several diagnostics to assess the quality of
 our simulations. 
In cases where several resolution simulations are considered and some solution is known, we monitor 
convergence of the solution by calculating the convergence order
$n_{\rm con}$ as
\begin{equation}\label{eq:convergence_rate}
n_{\rm con} = \dfrac{\log\left(\dfrac{L_i - L_0}{L_j - L_0}\right)}{\log\left(\Delta x_i/\Delta x_j\right)},
\end{equation}
where $L_0$ is the value that the solution should approach in the
continuum limit, $i$ is a label for a grid of a 
given resolution, $L_i$ corresponds to the solution  on grid $i$, and 
$\Delta x_i$ corresponds to the grid 
resolution for grid $i$. For all cases involving fluid variables we 
typically consider the reference solution $L_0$ in 
Eq.~\eqref{eq:convergence_rate} to be the initial data itself. 
The use of such a reference solution can prove troublesome in cases with 
unreliable or faulty initial data.
In several of the cases we consider, we find a lack of convergence toward 
the continuum solution even at the start of the simulation, suggesting 
non-convergent initial data.
In such cases we consider
the \textit{self}-convergence of a given solution, which is 
demonstrated when
\begin{equation}\label{eq:self_convergence}
\begin{aligned}
(L_{\rm CR} - L_{\rm HR}) \left[ {\left(\dfrac{\Delta x_{\rm MR}}{\Delta x_{\rm HR}}\right)}^n  - 1  \right] =\\
(L_{\rm MR} - L_{\rm HR}) \left[ {\left(\dfrac{\Delta x_{\rm CR}}{\Delta x_{\rm HR}}\right)}^n - 1  \right],
\end{aligned}
\end{equation}
where $n$ is the expected convergence order, $L$ is the solution on a given grid, $\Delta x$ is the grid 
resolution for a given grid, and the labels CR, MR, and HR correspond to the canonical, medium, and high 
resolution grids, respectively. We refer to quantities scaled according to Eq.~\eqref{eq:self_convergence} using 
the symbol $c[L]$, for quantity $L$. 

We monitor the maximum rest mass 
density $\rho_{\rm b, max}$ and minimum of the lapse function 
$\alpha_{\rm min}$ to qualitatively assess stability and collapse. 
We also consider spacetime quantities, such as the Hamiltonian constraint 
violations, in order to assess convergence of the solution.
We consider 2D snapshots of several fluid quantities, including the rest 
mass density $\rho_{\rm b}$, temperature $T$, and different components of 
the magnetic field strength $B_i$, where relevant. When considering 
rotational properties of the fluid, we compute the angular velocity in 
the equatorial plane as
\begin{equation}\label{eq:angvel}
\Omega = \dfrac{Xv^y - Yv^x}{\varpi^2},
\end{equation}
where $\varpi \equiv \sqrt{X^2 + Y^2}$, $X\equiv (x - x_{\rm com})$, 
$Y\equiv (y - y_{\rm com})$, $x$ and $y$ are 
the Cartesian coordinates, and $x_{\rm com}$ and $y_{\rm com}$ correspond 
to the coordinates of the Newtonian center-of-mass.

\begin{figure*}[htb]
\centering
\includegraphics{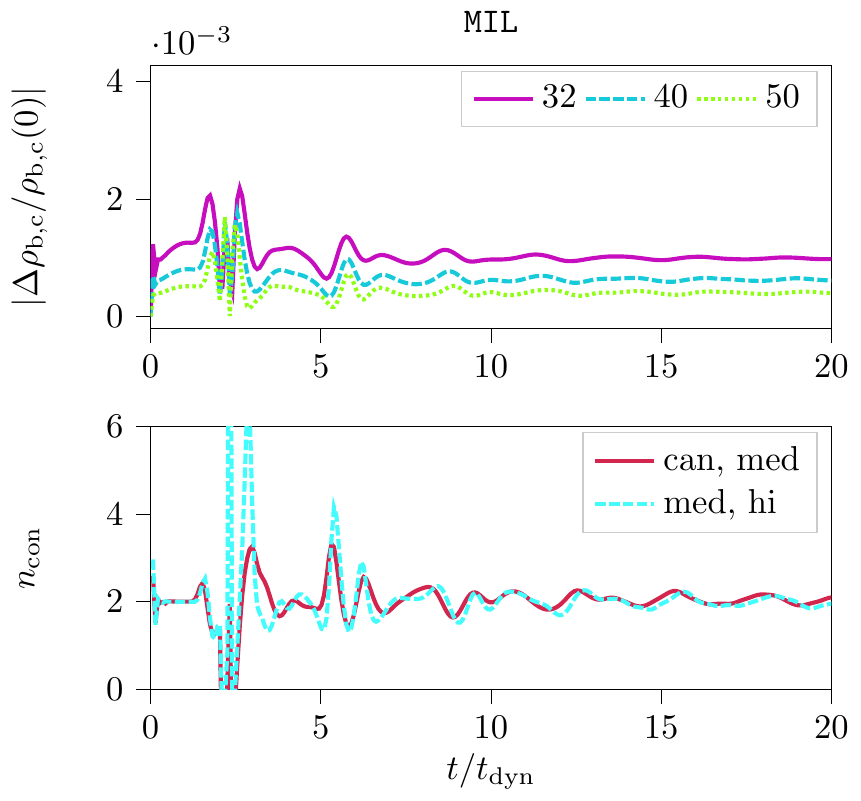}
\includegraphics{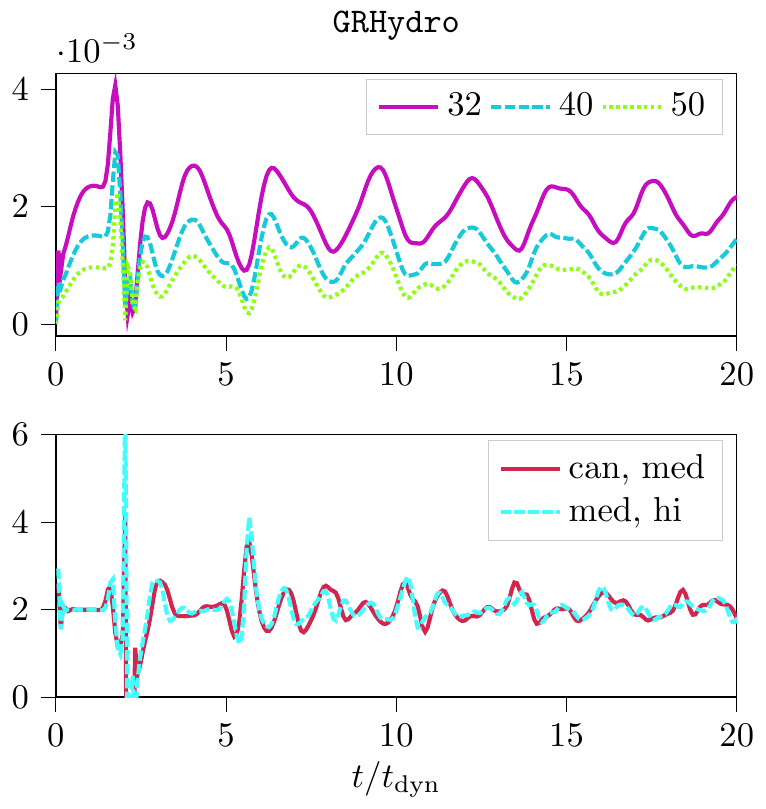}
\caption{\textit{Left panel}: Results of test ${\rm G2TOV}$ in the case
of evolution with the \texttt{MIL} code. 
In the top panel we show the relative change in central rest mass density 
compared to the value at 
$t=0$, where 
$\Delta \rho_{\rm b,c} \equiv \rho_{\rm b,c}(t) - \rho_{\rm b,c}(0)$. We 
depict results for grid 
resolutions 
that employ 32, 40, and 50 grid points per NS radius using solid magenta, dashed blue, and dotted green lines, 
respectively. In the bottom panel we show the convergence rate, calculated using Eq.~\eqref{eq:convergence_rate}, 
for the quantity depicted in the top panel in cases where we compare 
the canonical/medium and medium/high resolution 
results using solid red and dashed blue lines, respectively. \textit{Right panel}: same as left panels but for 
the \texttt{GRHydro} code.
}%
\label{fig:convergence_TOVG2}
\end{figure*}

In cases where we consider gravitational radiation, we extract GWs 
using the Newman-Penrose formalism~\cite{Newman:1961qr, Penrose:1962ij}, 
with focus on the $s=-2$ spin-weighted spherical harmonic
decompositions of the Newman-Penrose scalar $\Psi_4$. The coefficients of 
the spin-weighted decomposition are 
labeled $\Psi_4^{l,m}$, where $l$ and $m$ are the degree and order for 
the spherical harmonics. 
We extract $\Psi_4^{l,m}$ from the numerical solution at fixed concentric 
spheres with increasing coordinate 
radii
$r_{\rm ex} = \eta~\SI{}{km}$, where $\eta$ takes on several discrete 
values $70 \lesssim \eta \lesssim 450$. When extracting GWs, we ensure to 
use the values of $\Psi_4^{l,m}$ which are in the wave zone.
Where relevant, we compute the GW strain $h$ from
\begin{equation}
\Psi_4 = \ddot h_+ - i \ddot h_\times.
\end{equation}
using the fixed-frequency integration (FFI) 
method~\cite{Reisswig_2011GWs}. We
identify the time of merger $t_{\rm mer}$ as the time when the GW 
amplitude reaches an absolute maximum during the merger.

\subsection{G2TOV\@: Test using a TOV star with polytropic EOS}%
\label{subsec:G2TOV}
We first consider the case of TOV initial data constructed using a 
$\Gamma=2$ polytropic 
EOS\@. We construct initial data using the \texttt{TOVSolver} thorn within
the \texttt{EinsteinToolkit} with a 
central rest mass density of $\rho_{\rm b,c} = 0.129$ and polytropic 
constant of $\kappa=1$. We consider a static 
spacetime and do not employ mesh refinement (i.e., the test is carried 
out in the Cowling approximation). This is the only test we consider 
under the Cowling approximation. The half-side length of the 
solution grid extends to 
$1.25 R_{\rm NS}$, where $R_{\rm NS}\approx \SI{8}{M_\odot}$ is the NS 
radius. Our canonical, medium, and high 
resolution grids employ grid spacings such that there are 32, 40, and 50 
grid 
points covering the NS radius, respectively. We evolve with the 
\texttt{MIL} code
and with the \texttt{GRHydro} code as a 
comparison. In the case of the \texttt{GRHydro} evolution 
we employ the Harten-Lax-vanLeer-Einfeldt 
(HLLE)~\cite{doi:10.1137/0725021} Riemann solver and 
PPM reconstruction for maximum algorithmic overlap with \texttt{MIL}.

In the top panel of Fig.~\ref{fig:convergence_TOVG2} we show the evolution 
of the relative change in the 
central rest mass density, where 
$\Delta \rho_{\rm b,c} \equiv \rho_{\rm b,c}(t) - \rho_{\rm b,c}(0)$
over a timescale of $20\, t_{\rm dyn}$, where
\begin{equation}\label{eq:dynamical_time}
t_{\rm dyn} = 1/\sqrt{\rho_{\rm b,c}(0)}
\end{equation}
is the dynamical time. 
We show results for the low, medium, and high resolution grids, using 
solid magenta, dashed blue, and dotted green lines, respectively. We find 
that the evolution of a polytropic TOV 
star proceeds as expected, with oscillations in the rest mass density not 
exceeding $\sim 0.2\%$  and 
decreasing over time as the model settles. In the lower panel
of Fig.~\ref{fig:convergence_TOVG2}
we also depict the convergence order of 
$\Delta \rho_{\rm b,c}/\rho_{\rm b,c}(0)$, 
with comparisons of the canonical/medium and medium/high resolution results
depicted with the solid red and dashed blue lines, respectively. As we 
increase the grid resolution we find that 
the quantity $\Delta \rho_{\rm b,c}$ converges 
to zero at the expected second-order rate. In the right panel of 
Fig.~\ref{fig:convergence_TOVG2} we 
show the results of the ${\rm G2TOV}$ test in the case of evolution with 
the \texttt{GRHydro} code. We find 
behavior in the \texttt{GRHydro} code which is consistent with that of 
the \texttt{MIL} code. When evolving 
with \texttt{GRHydro} we find that oscillations in the rest mass density 
are of larger amplitude when compared to 
the \texttt{MIL} code, but the solution nonetheless converges at the 
expected rate. 

\subsection{LSTOV\@: Test using a cold, magnetized  TOV star with realistic EOS}%
\label{subsec:realTOVCook}
\begin{figure*}[htb]
\centering
\includegraphics{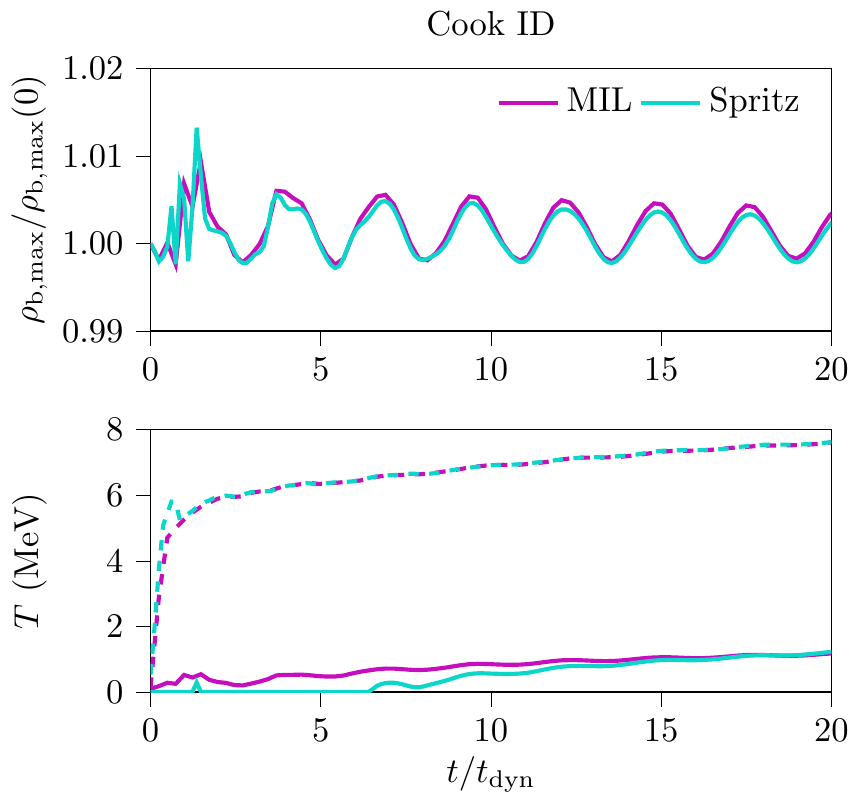}
\includegraphics{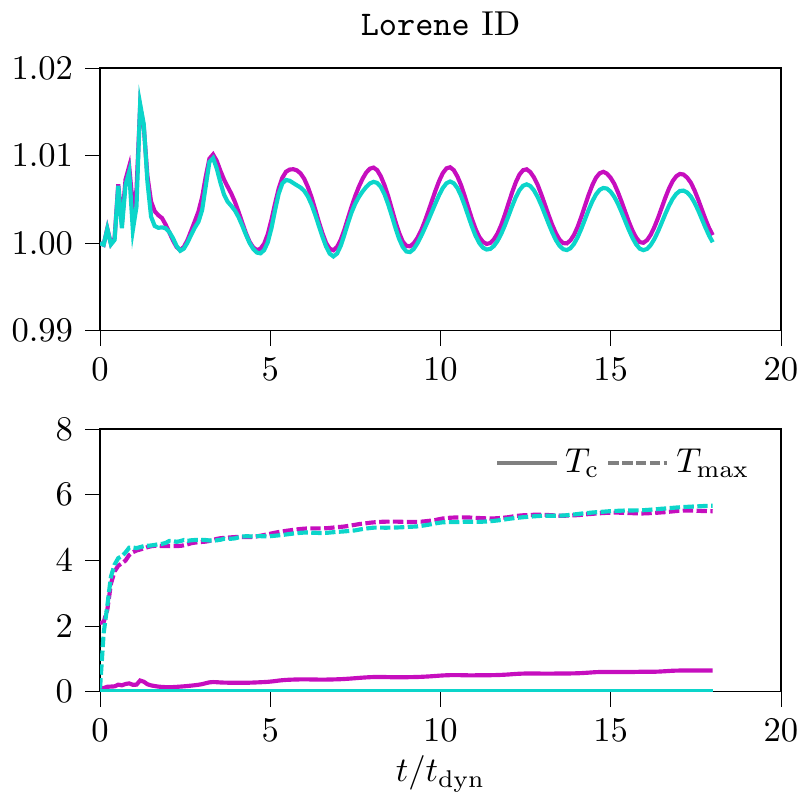}
\caption{\textit{Left panel:} Results of test ${\rm LSTOV}$ in the case
of evolution with the \texttt{MIL} code. We depict the central rest mass density $\rho_{\rm b,c}$ scaled by the value at $t=0$ (top panel), 
central temperature $T_{\rm c}$ (lower panel, solid lines), and maximum temperature $T_{\rm max}$ (lower panel, 
dashed lines) as functions of time (scaled by the dynamical time) for a magnetized TOV star built with the LS220 
EOS using the Cook code~\cite{CST94a}. We show the results corresponding 
to evolution with the {\tt MIL} (magenta lines) and the \texttt{Spritz} 
(blue lines) codes. \textit{Right panel:} Same as the left 
panel, but in the case where initial data was built using 
the \texttt{Lorene} code~\cite{LORENE}. }%
\label{fig:evolution_TOVLS}
\end{figure*}

In this set of tests we consider a cold, magnetized TOV star built using a realistic EOS\@.
We construct initial data with several codes as a way to understand the systematic error introduced at the level 
of the initial data. 
In particular, we use the Cook code~\cite{CST94a}, the \texttt{RNSID} code~\cite{Stergioulas1995, Nozawa1998} 
packaged within the
\texttt{EinsteinToolkit}, and the \texttt{Lorene} code~\cite{LORENE}. In the following we focus on initial data constructed with the Cook and {\tt Lorene} codes, as the {\tt RNSID} code produces similar results. We construct 
initial data for a cold, barotropic, $\beta$-equilibrated EOS obtained using the LS220 EOS\@. The initial configuration has
a central energy (rest mass) density of $\SI{0.8e15}{\g\per\cm\cubed}$ $(\SI{0.74e15}{\g\per\cm\cubed})$,
ADM (rest) mass of $\SI{1.4}{M_\odot} (\SI{1.55}{M_\odot})$, and circumferential radius of 
$R_{\rm NS} = \SI{12.69}{km}$. We employ 3 levels of fixed mesh refinement, with the innermost, finest refinement 
level extending to $1.5 R_{\rm NS}$, and every subsequent level extending to twice the distance of the
adjacent finer level. 
For all tests we employ a finest-level grid 
resolution of $dx_{\rm fin}=R_{\rm NS}/64$ (which we refer to as 
canonical resolution), such that there are 64 grid points covering the NS
radius. We superimpose a purely poloidal magnetic on non-magnetized initial 
data which is confined to the interior of the NS, using the {\tt 
Seed\_Magnetic\_Fields\_BNS} code within the {\tt EinsteinToolkit}. The field structure approximately 
corresponds to that generated by a current loop~\cite{Paschalidis:2013jsa}, with the toroidal component of the 
vector potential taking the form 
\begin{equation}\label{eq:A_phi}
A_\phi = \dfrac{\pi r_0^2 I_0 \varpi^2}{{(r_0^2 + r^2)}^{3/2}} \left(1 + \dfrac{15r_0^2(r_0^2 + \varpi^2)}{8{(r_0^2 + r^2)}^2} \right),
\end{equation}
where $r_0$ is the current loop radius, $I_0$ is the loop current, 
$r^2 = \varpi + z^2$, $\varpi = {(x-x_{\rm NS})}^2 + {(y-y_{\rm NS})}^2$, and
$x_{\rm NS}$ and 
$y_{\rm NS}$ are the initial coordinates of the NS center-of-mass. This 
choice of magnetic field exhibits a 
$1/r^3$ fall-off outside the NS, as expected for a dipole field, but as 
mentioned above we do not extend the field to the exterior; the {\tt 
Seed\_Magnetic\_Fields\_BNS} code employs a cut-off pressure $P_{\rm cut}$ 
below which the vector potential is set to zero, which we set to 
$P_{\rm cut}\approx0.15 P_{\rm max}(0)$ (where $P_{\rm max}(0)$ is the 
maximum fluid pressure at the start of the simulation). 
The initial field strength corresponds to a ratio 
of magnetic to fluid pressure of $P_{\rm mag}/P \approx 0.001$.

As a comparison, we also show 
the evolution of this model with the \texttt{Spritz} 
code~\cite{2020zndo...3689751G}, which uses
an implementation of the \texttt{Palenzuela} solver for finite 
temperature, tabulated EOSs. We employ the \texttt{Spritz} code with 
PPM reconstruction and the HLLE Riemann solver for maximum algorithmic 
overlap with \texttt{IllinoisGRMHD}. In Fig.~\ref{fig:evolution_TOVLS} we 
show the evolution of the central rest mass density $\rho_{\rm b, c}$ 
(top panels), maximum temperature $T_{\rm max}$ (bottom panels using 
dashed lines), and central temperature $T_c$
(bottom panels using solid lines) as functions of time. 
In the left panel of 
Fig.~\ref{fig:evolution_TOVLS} we focus on cases where initial data was 
constructed with the Cook code. We depict results corresponding to 
\texttt{MIL} and \texttt{Spritz} using magenta and blue 
lines, respectively.
We find that the evolution of the rest mass density
is qualitatively the same between the \texttt{MIL} the \texttt{Spritz} 
codes, exhibiting oscillations of at most $\sim 1\%$. 
The configuration shows an initial relatively large oscillation in the 
rest mass density of $\gtrsim 1\%$ which peaks 
at $t\approx 1-2\,t_{\rm dyn}$.
This initial large oscillation is likely due to the creation of a cold 
barotropic EOS from the finite 
temperature LS220 EOS table as previously discussed, which requires 
interpolation and 
root-finding routines (we note that similar oscillations happens in test 
cases with unmagnetized initial configurations, which suggests that the 
superimposed magnetic field is likely not the culprit behind the 
oscillations). It is possible that the 
use of a cold EOS created from a finite temperature EOS
results in initial data with percent-level numerical errors, which is 
consistent with the oscillations of 
amplitude $\sim 1.01 \rho_{\rm b,c} (0)$ shown in the top left panel of 
Fig.~\ref{fig:evolution_TOVLS}. We note that initial data produced with 
the Cook code uses a compactified radial 
grid (true for both {\tt RNSID} and {\tt LORENE}) in spherical polar coordinates, and as such interpolation of the 
initial data onto
the Cartesian grids of the \texttt{Cactus} infrastructure results in 
additional numerical errors in the initial 
configuration. As the initial configuration settles to the nearest 
equilibrium, it begins to exhibit  
smaller oscillations in the rest mass density of $\sim 0.5\%$. These 
oscillations decay over a timescale of 
$t \approx 20\, t_{\rm dyn}$. In the bottom left panel of 
Fig.~\ref{fig:evolution_TOVLS}, 
we show the evolution of the maximum  and central temperatures using 
dashed 
and solid lines, respectively, for initial data constructed with the Cook 
code.
The initial configuration is cold for all solvers tested, with a central 
(maximum) temperature of 
$T_{\rm c} \approx \SI{0.1}{MeV}$ ($T_{\rm max} \approx \SI{0.5}{MeV}$). We 
note that initial data is constructed for an isothermal star at 
$T=\SI{0.1}{MeV}$. However, once the initial data for $\rho_b$, $\epsilon$, 
$Y_{\rm e}$, and $P$ are interpolated onto the solution grid, we recover 
the temperature for self-consistency among the fluid variables, which leads 
to errors in the temperature profile at $t=0$, near the stellar surface. 
Early in the simulations, regardless of the evolution code used, the model 
develops a warm atmosphere, which becomes increasingly warm as the star 
undergoes small oscillations. The
central configuration remains cold as the temperature of the atmosphere 
increases and saturates to 
$T_{\rm max}\approx \SI{8}{MeV}$. 

\begin{figure*}[htb]
\centering
\includegraphics{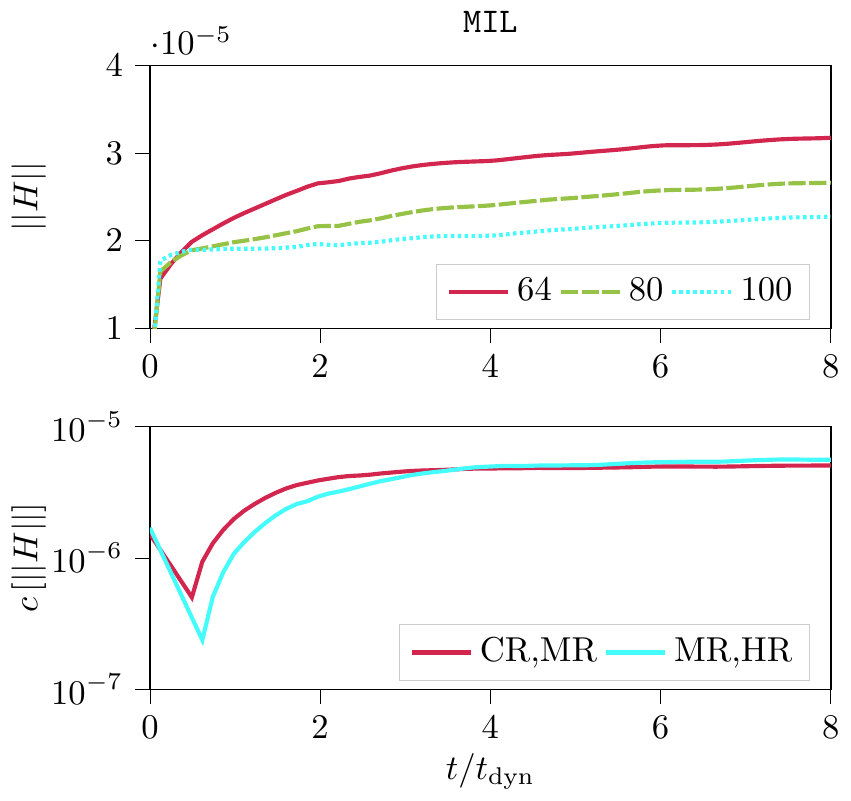}
\includegraphics{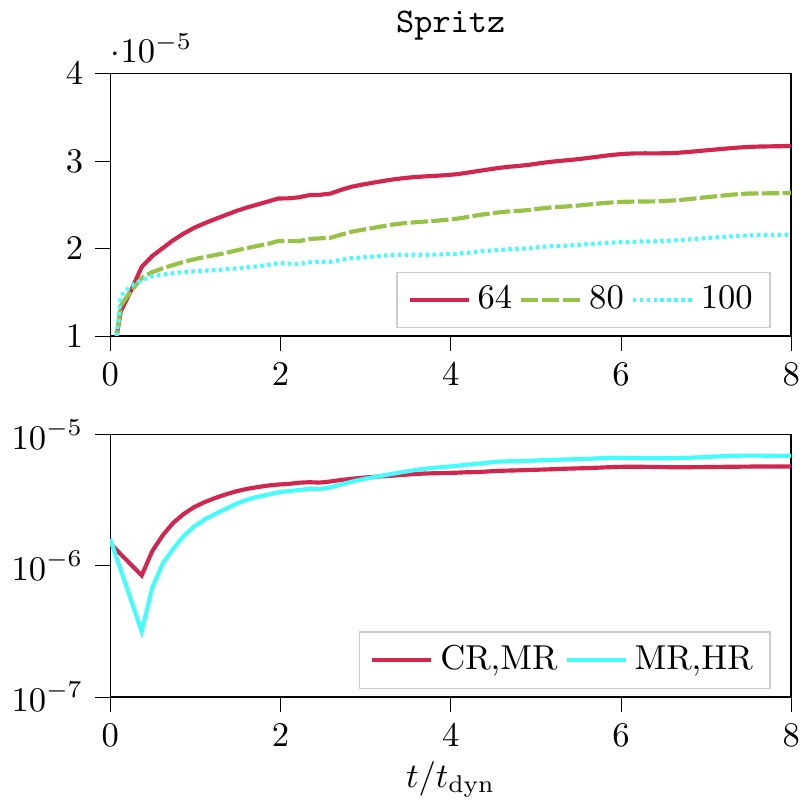}
\caption{\textit{Left panel:} Results of test ${\rm LSTOV}$ in the case
of evolution with the \texttt{MIL} code. We depict the convergence of the L2 norm of the Hamiltonian constraints $|| H ||$ in the case of a 
magnetized TOV star built using the LS220 EOS with the \texttt{Lorene} code, evolved with the \texttt{MIL} code 
and \texttt{Palenzuela} solver. The top panel depicts $||H||$ for the low, medium, and high resolution grids using 
solid red, dashed green, and dotted blue lines, respectively. 
The bottom panel depicts $||H||$ scaled according to second-order 
convergence while comparing the canonical and medium (CR,MR) and medium and high (MR,HR) resolution results using 
red and blue lines, respectively. 
\textit{Right panel}: Same as the left panel but in the case of evolution with the \texttt{Spritz} code. }%
\label{fig:convergence_TOVLS_LOR}
\end{figure*}

To test the role of initial data on the evolution of the model, 
we consider the same initial configuration but instead built 
with the \texttt{Lorene} and {\tt RNSID} codes. We consider the same 
central rest mass density and 
grid hierarchy as in the case considering initial data constructed with 
the Cook code. In the right panel of 
Fig.~\ref{fig:evolution_TOVLS} we present results of the evolution with 
the \texttt{MIL} and \texttt{Spritz} codes using magenta and blue lines,
respectively, with focus on the case of initial data built with the 
\texttt{Lorene} code.
In all cases considered, we observe initial oscillations of the rest mass 
density which do not exceed $2\%$ and subsequent 
oscillations which are at the level of $\sim 1\%$. We generally find that 
lower temperatures when using initial data constructed with {\tt Lorene}, 
in both the bulk of the star and in the atmosphere. We find that 
evolution with the \texttt{Spritz} code results in a central region which 
remains cold ($T\leq \SI{0.1}{MeV}$, shown using the blue solid line in 
the lower right panel of Fig.~\ref{fig:evolution_TOVLS}) 
throughout the simulation, whereas the \texttt{MIL} code produces 
slightly 
warmer central regions of up to $T_{\rm c}=\SI{0.3}{MeV}$. We note that in 
the \texttt{MIL} code we impose a lower bound on the temperature of 
$\SI{0.1}{MeV}$ to avoid EOS interpolations near the table limits. Such a 
low-temperature floor is not imposed in the \texttt{Spritz} code, which 
may be the reason for the slight 
difference in $T_{\rm c}$ between the two codes. A central 
temperature of $T_{\rm c}=\SI{0.3}{MeV}$ is still 
significantly cold from a nuclear perspective and tests of TOV star 
evolution without imposing the lower bound on 
$T$ results in a temperature profile which is consistent with the 
\texttt{Spritz} evolutions. 
The evolution of the rest mass density and temperature presented in 
Fig.~\ref{fig:evolution_TOVLS}
is typical of stable TOV stars. The fact that a realistic TOV star 
remains stable and cold over
a timescale of $t\approx20\, t_{\rm dyn}$ and the qualitatively similar
behavior between the \texttt{MIL} and 
\texttt{Spritz} codes indicates that our implementation of the solvers 
within \texttt{ConservativeToPrimitive} is reliable. Tests 
with {\tt RNSID} initial data produce very similar results to those 
presented in Fig.~\ref{fig:convergence_TOVLS_LOR}.

We consider the models constructed using the Cook code for a
resolution study. We evolve the Cook initial data in two additional
higher resolution simulations with both the \texttt{MIL} and
\texttt{Spritz} codes. For medium (high) resolution tests, we employ
grid resolutions of $dx_{\rm med}=dx_{\rm can}/1.25$ ($dx_{\rm
  hi}=dx_{\rm can}/1.25^2$) such that there are 80 (100) grid points
covering the NS radius, where $dx_{\rm can} = dx_{\rm fin}$ as listed
in Tab.~\ref{tab:grids_C2Ptests} for the LSTOV test.  In the top panel
of Fig.~\ref{fig:convergence_TOVLS_LOR} we show the L2 norm of the
Hamiltonian constraints $||H||$ for the canonical (solid red lines),
medium (dashed green lines) and high (dotted blue lines) resolution
grids, in cases where we evolve with the \texttt{MIL} (left panel) and
\texttt{Spritz} (right panel) codes. We find that $||H||$ is generally
small and converges toward zero (the expected value in the continuum
limit) for higher resolution grids. We also note that the evolution of
$||H||$ is in agreement between the \texttt{MIL} and \texttt{Spritz}
codes.  However, we find that $||H||$ converges toward zero at a rate
which is slower than expected, for both evolution codes
considered. Specifically, we find a linear convergence rate of $||H||$
toward zero. We note that the expected convergence rate of the
evolutions we consider is second-order, and that second-order
convergence is demonstrated for polytropic initial data (see
Fig.~\ref{fig:convergence_TOVG2}).  It is likely that the lower order
convergence observed for tests employing realistic EOSs is due to
errors in the initial data.  In particular, the initial data codes we
consider make use of compactified spherical polar coordinate grids,
which are not the same as the evolution grids. Numerical errors are
therefore inevitable when interpolating solutions from their original
compactified coordinates to the Cartesian grids employed within the
\texttt{EinsteinToolkit}.  We point out that the evolution of
magnetized TOV stars presented in this section is consistent with
stable evolution, and coincides with the results presented
in~\cite{2020zndo...3689751G} for \texttt{Spritz}. As there are likely
errors at the level of the initial data when using realistic EOSs we
instead consider the \textit{self}-convergence of the solution.  In
the lower panels of Fig.~\ref{fig:convergence_TOVLS_LOR} we show
$||H||$ scaled according to second order convergence (i.e., $n=2$ in
Eq.~\eqref{eq:self_convergence}) for test LSTOV while comparing the
canonical and high resolution (solid red lines) and medium and high
resolution solutions (dashed blue lines). We again show results for
evolution with the \texttt{MIL} and \texttt{Spritz} codes using the
left and right panels, respectively.  The lower panels of
Fig.~\ref{fig:convergence_TOVLS_LOR} demonstrate that the solution
exhibits self-convergence at the expected order. We also consider the
same test of convergence in cases with initial configurations built
with the \texttt{Lorene} and \texttt{RNSID} codes and find similar
results to those presented in Fig.~\ref{fig:convergence_TOVLS_LOR}. We
point out that the construction of TOV initial data is algorithmically
very similar between the Cook, \texttt{RNSID}, and \texttt{Lorene}
codes, as they all make use of compactified spherical coordinate
grids. As such, interpolation errors that results from a change in the
coordinate system will arise in all tested cases.

\subsection{G2BNS\@: Test of a BNS with a Gamma-law EOS}\label{subsec:G2BNS}
\begin{figure*}[htb]
\centering
\includegraphics{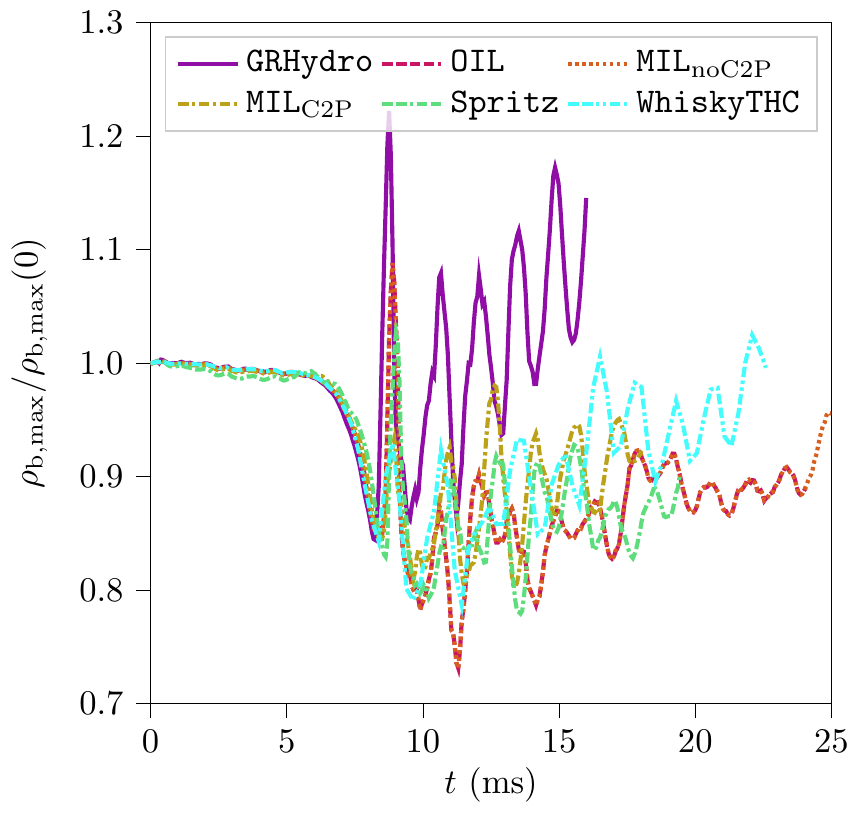}
\includegraphics{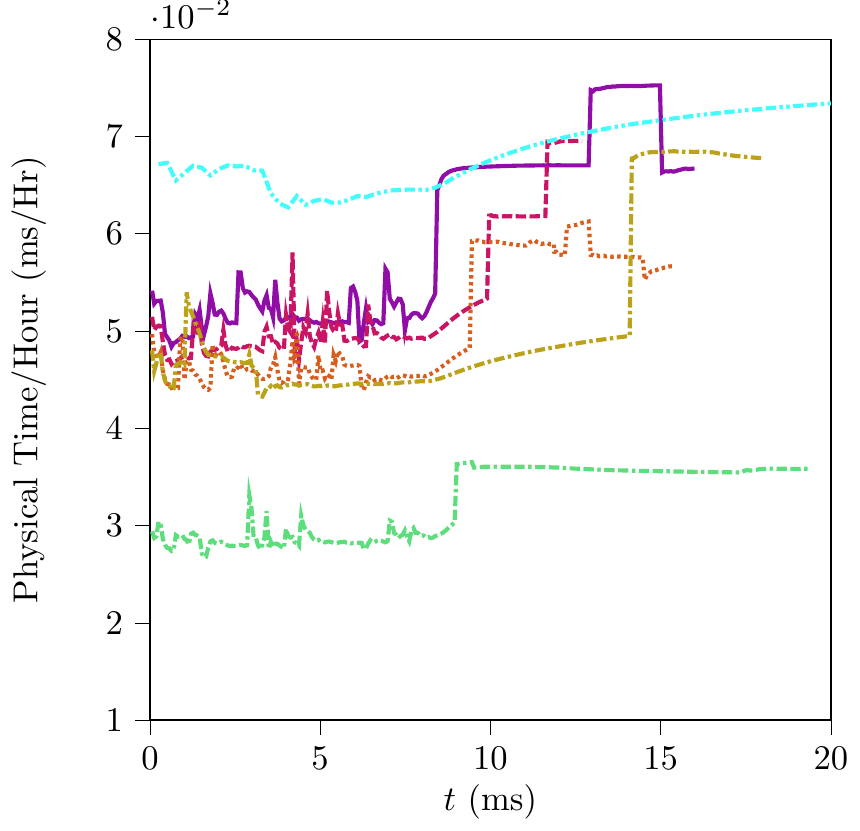}

\caption{\textit{Left panel:} Maximum rest mass density $\rho_{\rm b}$, 
scaled by the value at $t=0$, for an unmagnetized BNS assuming a 
$\Gamma=2$ polytropic EOS\@. We depict the results corresponding to the
{\tt GRHydro} (solid purple line), {\tt OIL} (dashed magenta line), 
{\tt MIL} with and without the use of the solvers within 
{\tt ConservativeToPrimitive} (dash-dotted golden and dotted orange 
lines, respectively), {\tt Spritz} (dash-dash-dotted line), 
and {\tt WhiskyTHC} 
(blue dash-dash-dot-dotted line) codes. 
\textit{Right panel:} Same format 
as the left panel, but instead depicting the physical time produced by 
each simulation per hour of run time, across 224 {\tt Intel Haswell V3} 
cores. We use identical initial data, grid structures, 
conservative-to-primitive error tolerance, and computational 
resources in all cases, as well as similar flux treatments and reconstruction schemes. 
The tests presented here are as close as possible to a one-to-one comparison 
between codes that we could achieve.
}%
\label{fig:G2BNS_unmag}
\end{figure*}

As an additional test of the solvers implemented within \texttt{ConservativeToPrimitive}, 
we consider the evolution of a BNS system built assuming a 
polytropic EOS\@. We consider this test because the \texttt{OIL} code
can reliably simulate such a system and because it provides a stringent test of the \texttt{MIL} code new algorithms in a highly 
relativistic scenario. We assume a polytropic constant of $\kappa=123.61$ and adiabatic
index of $\Gamma=2$. We construct initial data for an equal mass
BNS with an orbital separation of $\SI{45}{km}$ using the \texttt{Lorene} code; the ADM mass and radius of each 
binary component are $M_{\rm NS}=\SI{1.4}{M_\odot}$ and $R_{\rm NS} = \SI{10.36}{km}$, respectively. 
During the evolution, we assume a $\Gamma$-law EOS with adiabatic index
$\Gamma=2$. In this test we employ AMR, with two sets of nested boxes centered on each 
NS and each employing 7 levels of refinement. We employ an additional set of nested boxes with 7 levels of 
refinement centered at the origin of the solution to reliably resolve the BNS merger remnant. 
The outer grid boundary is 
located at $R_{\rm out}=\SI{582.12}{km}\approx 38.4 R_{\rm NS}$ and the half-side length of the finest level 
around each star extends to $\sim 1.25 R_{\rm NS}$ with each subsequent coarser level extending to 
$\eta R_{\rm NS}$ where $\eta \in (1.7,2.5,5.8,11.6,23.2)$. The finest-level grid resolution is set to 
$dx_{\rm fin} \approx R_{\rm NS}/40$, such that we initially resolve each 
binary component with at least
40 grid points along the radius. For test G2BNS, we consider both an 
unmagnetized, and magnetized 
scenario, because not all of the evolution codes considered can handle a 
constrained-transport evolution of the magnetic field. In cases where a 
magnetic field is also considered, we seed purely 
poloidal fields which are confined to the interior of each NS, with vector 
potential of the form 
Eq.~\eqref{eq:A_phi}, such that the initial maximum field 
strength is $B_{\rm max} \approx \SI{2.3e15}{G}$.

We simulate the system with the use of {\tt MIL}
and track the evolution of the rest mass density $\rho_{\rm b}$, fluid
3-velocity $v^i$, and the magnetic field $B^i$, when relevant. We 
approximate the merger time $t_{\rm mer}$ by
considering the time when the rest mass density reaches a maximum during 
the evolution. We compare to several other open-source codes in the 
literature, including the {\tt OIL}, {\tt GRHydro}, {\tt WhiskyTHC}, and 
{\tt Spritz} codes. Similar to other tests, we employ all evolution codes 
considered with the HLL/HLLE Riemann solver. For the {\tt IllinoisGRMHD}, 
{\tt GRHydro}, and {\tt Spritz} cases we use PPM reconstruction. For the 
{\tt WhiskyTHC} case we use MP5 reconstruction~\cite{SURESH199783}. We also note that 
{\tt WhiskyTHC} uses an internal rescaling of the rest mass density, and as such the 
initial conditions between {\tt WhiskyTHC} and other cases is not identical. However, 
this internal rescaling of the rest mass density results in a maximal relative 
difference of less than $0.4\%$ between cases.

\subsubsection{Unmagnetized BNS}

In cases where we do not consider the evolution of the magnetic field, we 
are free to compare the results provided by all of the open-source GRHD 
codes considered, as they are all able to handle an unmagnetized BNS 
system built with a polytropic EOS\@. As such, we compare the results of
simulations using the {\tt MIL}, {\tt OIL}, {\tt GRHydro}, 
{\tt WhiskyTHC}, and {\tt Spritz} codes. We emphasize that the numerical 
grids, initial data, reconstruction scheme, conservative-to-primitive schemes, and 
flux treatment for all of 
these simulations are chosen for maximum overlap, with the latter three likely having 
differences in implementation between codes. The comparison of results 
presented in this section was the closest scenario to a one-to-one 
comparison between open-source numerical codes that we could obtain.

\begin{figure*}[htb]
\centering
\includegraphics[width=18cm]{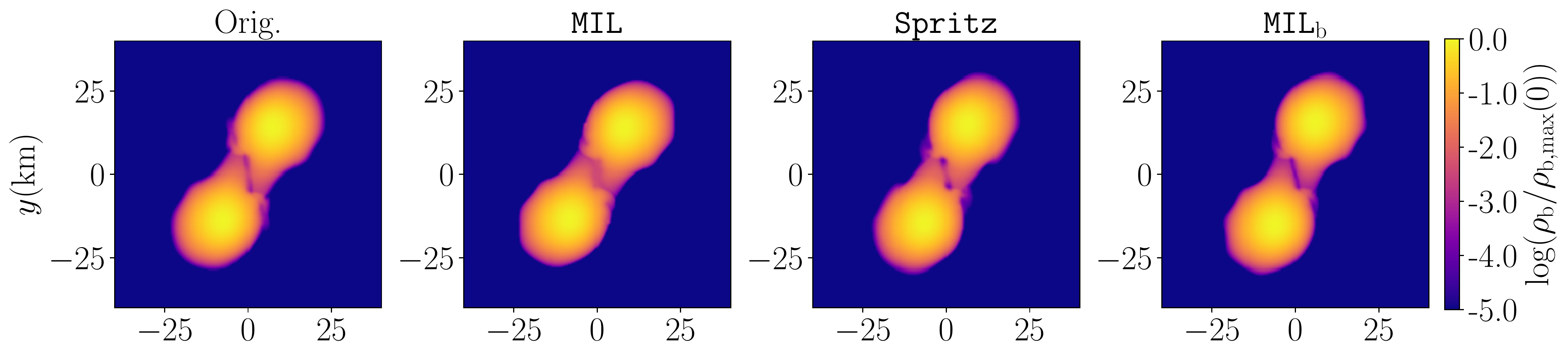}
\includegraphics[width=18cm]{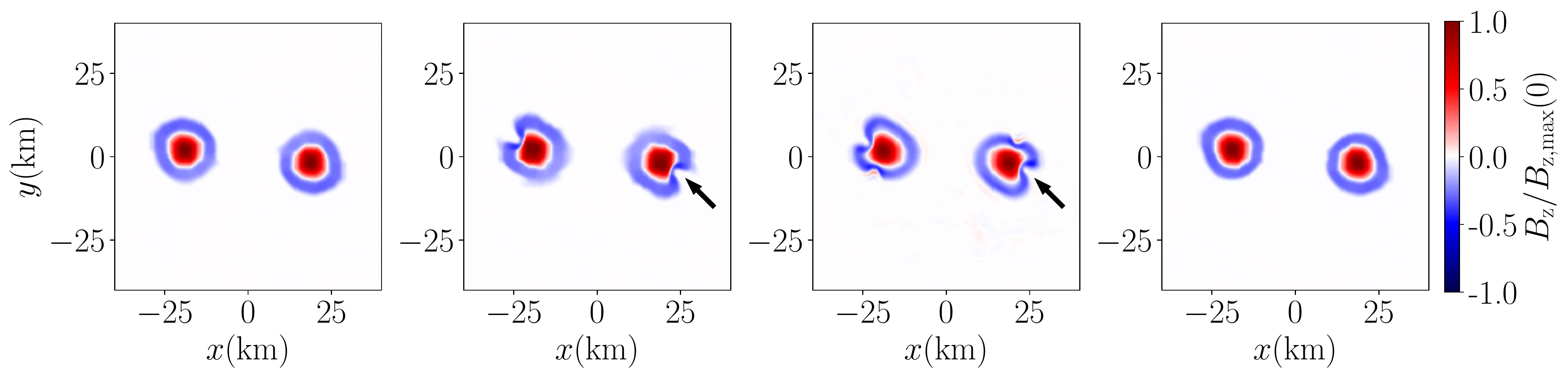}
\caption{\textit{Top panel}: Equatorial snapshots of the rest mass 
density $\rho_{\rm b}$ 
(scaled by the maximum value at $t=0$) 
for test G2BNS at $t\approx\SI{7.3}{ms}$.
From left to right we show results for the \texttt{OIL} code, 
\texttt{MIL} code with \texttt{Palenzuela} solver, and 
\texttt{Spritz} code. The rightmost panel depicts a snapshot for an 
evolution using the 
{\tt MIL} code wherein we use an alternative calculation of the specific 
enthalpy $h$ (see Eqs.~\eqref{eq:enthalpy_original} 
and~\eqref{eq:enthalpy_c2p}).
\textit{Bottom panel}: Equatorial snapshots of the $z-$component of the 
magnetic field $B_z$ (scaled by the 
maximum value at $t=0$) for test ${\rm G2BNS_{mag}}$ at 
$t\approx\SI{3.54}{ms}$ for the same set of codes as the top panel.
We point out common features that develop in the magnetic field structure 
using black arrows.
}%
\label{fig:test_c2p_G2BNS}
\end{figure*}

In the left panel of Fig.~\ref{fig:G2BNS_unmag} we show the maximum rest 
mass density in the 
simulations for the unmagnetized case of test G2BNS\@. We show results for
the {\tt GRHydro} (solid purple line), {\tt OIL} (dashed magenta line), 
{\tt MIL} with and without\footnote{We consider an evolution with the 
{\tt MIL} code that does not employ the solvers within {\tt 
ConservativeToPrimitive}, and instead uses the original solvers within 
the {\tt OIL} code. We label this test as ${\tt MIL_{\rm noC2P}}$ in 
Fig.~\ref{fig:G2BNS_unmag}. The fact that the ${\tt MIL_{\rm noC2P}}$ 
evolution proceeds identically to the {\tt OIL} case ensures that other 
changes made to the code (e.g., the changes that allow for passive 
$Y_{\rm e}$ advection) do not affect the dynamics.} the use of the solvers within 
{\tt ConservativeToPrimitive} (dash-dotted golden and dotted orange 
lines, respectively), {\tt Spritz} (dash-dash-dotted line), 
and {\tt WhiskyTHC} (blue dash-dash-dot-dotted line) codes. We find that 
during the inspiral, 
all codes behave qualitatively the same, with the maximum rest mass 
density exhibiting oscillations of $\lesssim 2\%$ for the first 
$t\approx 5\text{ ms}$ of the inspiral. We find that the merger time for 
these simulations all fall within $9.81 \pm 0.1 \text{ ms}$, resulting in 
a difference of at most $2\%$. On 
the other hand, we find significant 
deviations of the simulation results after the merger. Each code 
considered results in different size oscillations of the rest mass 
density for the first $10 \text{ ms}$ after merger. The {\tt GRHydro} 
code results in the largest post-merger oscillations in 
$\rho_{\rm b, max}$, and is the only case where $\rho_{\rm b, max}$ shows 
a tendency to grow. In all cases besides {\tt GRHydro}, the evolution of 
$\rho_{\rm b, max}$ is qualitatively similar, and the post-merger value 
of $\rho_{\rm b, max}$ oscillates between 
$0.9\rho_{\rm b, max}(0)-\rho_{\rm b, max}(0)$. We find qualitatively
similar GW emission in all cases, but the frequency corresponding to 
peak GW emission shows a spread of at most $\sim 15\%$ between cases.
Given that all codes 
use similar flux treatments (we note that in all other tested codes, the 
closest Riemann solver to the HLL solver adopted in {\tt IllinoisGRMHD} 
is the positivity-preserving HLLE solver~\cite{doi:10.1137/0725021}, 
which is what we adopt in evolutions with {\tt Spritz}, {\tt GRHydro}, 
and {\tt WhiskyTHC}) and reconstruction schemes (in the case of {\tt WhiskyTHC}, we 
use 
MP5 while in all other cases PPM is used), along with identical
numerical grid, initial data, conservative-to-primitive error tolerance, 
and spacetime evolution scheme, it is 
unclear why the {\tt GRHydro} 
code deviates so strikingly from all other codes, and 
why there is a 
general lack of quantitative agreement in the post-merger evolution 
between most codes. 
Key differences between codes may arise in the implementation of 
conservative-to-primitive algorithms, which may be the chief reason for 
discrepancies between codes. We leave the investigation of algorithmic 
implementations within open-source GRHD codes, and a more detailed code 
comparison, to future work.

In the right panel of Fig.~\ref{fig:G2BNS_unmag} we show a measure of the 
performance for each code. Specifically, we show the amount of physical 
time produced per hour for each simulation, using the same amount of 
computational resources (specifically, 224 {\tt Intel Haswell V3} 
cores). We find that, for this case study, the 
{\tt WhiskyTHC} code performs the fastest (producing around 0.07 ms/hour  
throughout the inspiral and in the post-merger phase, respectively), 
while the {\tt Spritz} code performs the slowest (producing around 0.03 
ms/hour during the inspiral and 0.035 ms/hour in the post-merger phase, 
respectively). All other codes perform similarly, with 0.05 ms/hour 
during the inspiral and 0.07 ms/hour in the post-merger. We note that the 
hierarchy suggested by the right panel of Fig.~\ref{fig:G2BNS_unmag} is 
not a final verdict on the performance of each code, and only 
demonstrates the performance for the particular case considered. We find 
that changes to the specific settings governing the 
conservative-to-primitive inversion and atmosphere control can 
significantly impact the performance of the codes considered. For 
example, we find that the conservative-to-primitive routines within 
{\tt Spritz} often perform as reliably as in other codes with the use of 
fewer algorithmic iterations, which serves to speed up the code.

\subsubsection{Magnetized BNS}

In cases where we \emph{do} consider a magnetic field for test G2BNS, 
we compare the evolution provided by the {\tt MIL} code to 
that provided by the
\texttt{OIL} and \texttt{Spritz} codes (note that the {\tt Spritz} code 
uses an implementation of the \texttt{Noble} solver for polytropic EOSs). 
In the top panel of Fig.~\ref{fig:test_c2p_G2BNS} we show equatorial 
snapshots of the rest mass density for test 
G2BNS at time $t\approx\SI{7.3}{ms}$, 
corresponding to first contact between the binary 
components. 
From left to right in Fig.~\ref{fig:test_c2p_G2BNS} we show snapshots for 
evolution with the \texttt{OIL}, \texttt{MIL}, and 
\texttt{Spritz} codes. In the rightmost panel of 
Fig.~\ref{fig:test_c2p_G2BNS} we also show a snapshot 
corresponding to the {\tt MIL} code in the case where we use an 
alternative set of reconstructed variables to calculate the specific 
enthalpy $h$ (labeled $\texttt{MIL}_{\tt b}$), as will be discussed in 
further detail below. We find that all codes 
considered produce very similar results as far as the merger time
and features during inspiral are concerned. 
We find that the merger times agree to within 2\%, specifically $t_{\rm mer}=47.85 t_{\rm dyn}$, 
$t_{\rm mer}=47.85 t_{\rm dyn}$, and $t_{\rm mer}=49.04 t_{\rm dyn}$ for 
the \texttt{OIL}, \texttt{MIL} and \texttt{Spritz} cases, 
respectively. At a time close to and preceding merger, pictured in the 
top panel of Fig.~\ref{fig:test_c2p_G2BNS}, the main
difference between the results of each code can be observed in the region 
between the binary components, likely due to how low density, 
near-atmosphere regions are treated in the conservative-to-primitive 
solvers implemented in each code.

In the lower panel of Fig.~\ref{fig:test_c2p_G2BNS} we show snapshots of 
the $z-$component of the magnetic field 
during the inspiral at time $t\approx\SI{3.54}{ms}$. Moving from left to 
right, we again show results for 
the \texttt{OIL}, \texttt{MIL}, and \texttt{Spritz} codes; the rightmost 
panel corresponds to the {\tt MIL} code with primitive 
reconstruction that agrees with the {\tt OIL} code 
(see Eq.~\eqref{eq:enthalpy_original}), which we label 
$\texttt{MIL}_{\tt b}$. Specifically, 
in \texttt{OIL} the calculation of specific enthalpy in the flux 
computation uses an analytic expression for $\epsilon$, based on 
a $\Gamma-$law EOS\@.
In the case where the adiabatic index for the thermal component of 
the EOS is $\Gamma_{\rm th}=2$, the specific enthalpy becomes
\begin{equation}\label{eq:enthalpy_original}
h = 1 + \epsilon(P_{\rm r}) + \dfrac{P_{\rm r}}{\rho_{\rm b,r}} = 1 + \dfrac{2P_{\rm r}}{\rho_{\rm b,r}},
\end{equation}
where $P_{\rm r}$ and $\rho_{\rm b,r}$ are the reconstructed pressure and 
rest mass density, respectively. On the
other hand, the solution within \texttt{MIL} and 
the \texttt{Spritz} codes calculate these 
specific enthalpies based on the reconstructed values of $\epsilon$
\begin{equation}\label{eq:enthalpy_c2p}
h = 1 + \epsilon_{\rm r} + \dfrac{P(\epsilon_{\rm r}, Y_{\rm e,r}, \rho_{\rm b,r})}{\rho_{\rm b,r}},
\end{equation}
where $P(\epsilon_{\rm r}, Y_{\rm e,r}, \rho_{\rm b,r})$ is the pressure 
for a generic EOS based on the 
reconstructed values of the specific internal energy $\epsilon_{\rm r}$, electron 
fraction $Y_{\rm e,r}$, and rest mass density $\rho_{\rm b,r}$. In other 
words, the \texttt{OIL} code uses the reconstructed pressure to calculate 
$h$, whereas the \texttt{MIL} and \texttt{Spritz} codes use the 
reconstructed specific internal energy.
The calculation of $h$ based on Eq.~\eqref{eq:enthalpy_c2p} is 
required for self-consistency among the fluid variables in the case of 
generic EOSs, but in the case of analytic 
EOSs, it is preferable to proceed with calculation of $h$ based on 
Eq.~\eqref{eq:enthalpy_original} for accuracy. The differences in use of 
reconstructed variables plays a role in the preservation of the magnetic 
field structure.
We find that the \texttt{OIL} code tends to advect the magnetic field 
structure accurately for longer times when 
compared to the other tested codes. The \texttt{MIL} and \texttt{Spritz} 
evolutions all 
produce similar features in the magnetic field near the low density parts 
of the star. In particular, solutions 
employing generic solvers, which are compatible 
with realistic EOSs, produce `kinks' in the magnetic field structure, as 
highlighted by the arrows in the lower 
panel of Fig.~\ref{fig:test_c2p_G2BNS}. 
We note that, despite the artifacts seen in the magnetic field structure 
for solvers compatible with realistic EOSs, the magnetic field is
generally properly advected during inspiral in all cases. Moreover,
the sizes of the artifacts decrease (and the time at which they appear
increases) with the use of higher resolutions grids.  The agreement of
\texttt{MIL} with the well-tested \texttt{Spritz} and \texttt{OIL}
codes suggests that the fluid and magnetic field solutions these codes
provide are reliable for highly dynamical systems.  For the sake of
consistency with the {\tt OIL} code, we adopt the same methods within
{\tt MIL} for calculating specific enthalpies in cases where we
consider analytic EOSs. We find that this results in a field structure
that is conserved throughout more of the inspiral, as illustrated by
the disappearance of the `kinks' in the field structure in the
rightmost snapshot in Fig.~\ref{fig:test_c2p_G2BNS}, wherein we use
the same method for calculating specific enthalpies as in 
the {\tt OIL} code. 
In other words, moving forward, within the {\tt MIL} code we calculate
$h$ based on Eqs.~\eqref{eq:enthalpy_original} and
Eq.~\eqref{eq:enthalpy_c2p} for analytic and finite temperature EOSs,
respectively.

\subsection{LSBNS\@: Tests using a BNS with finite temperature equations of state}%
\label{subsec:long_term}
\begin{figure}[htb]
\centering
\includegraphics{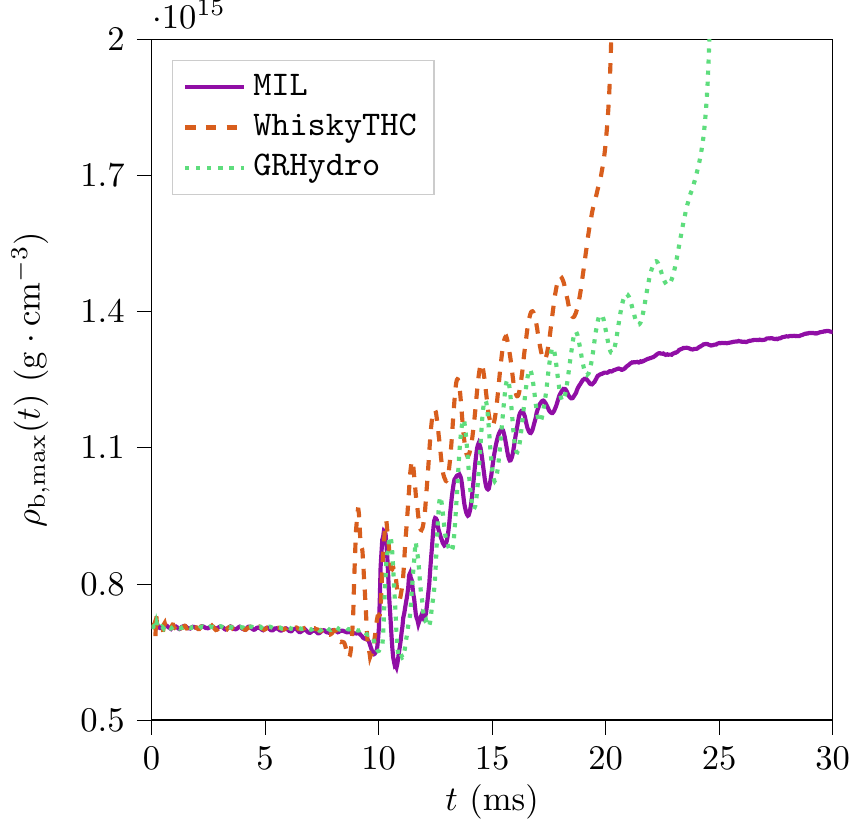}
\caption{Maximum rest mass density $\rho_{\rm b, max}$ as a function of 
time for the inspiral and merger of a BNS using the LS220 EOS (test 
LSBNS). We depict the evolution of $\rho_{\rm b, max}$ in the case of the 
{\tt MIL}, {\tt WhiskyTHC}, and {\tt GRHydro} codes using solid purple, 
dashed orange, and dotted green lines, respectively.}%
\label{fig:rho_LSBNS}
\end{figure}

\begin{figure}[htb]
\centering
\includegraphics{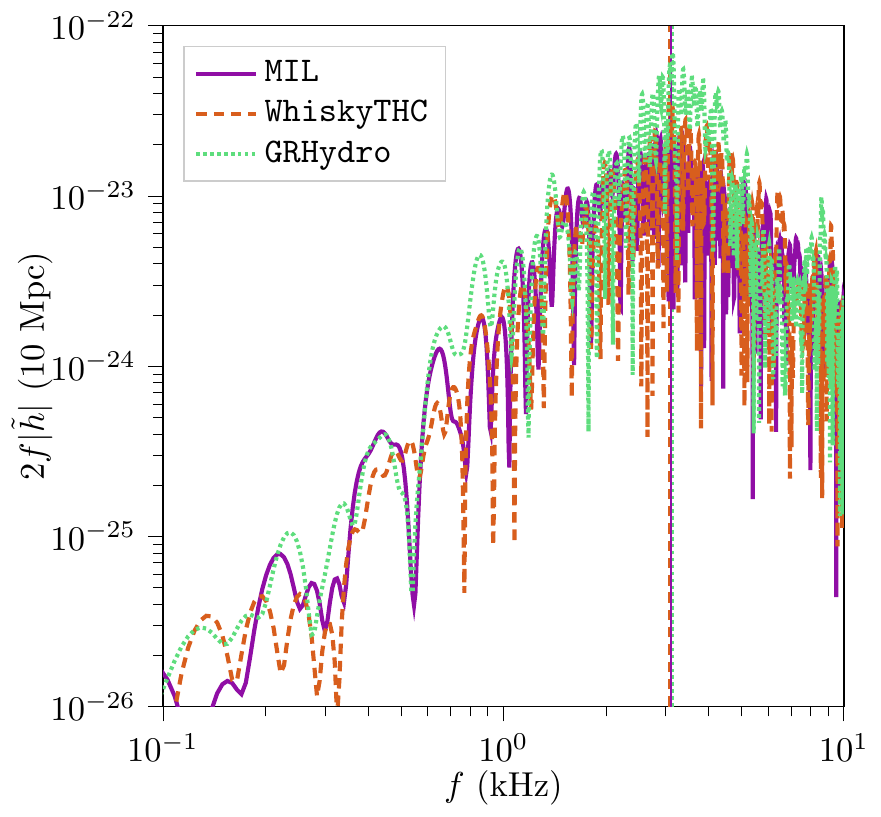}
\caption{GW signals corresponding to test LSBNS\@. We depict the
characteristic strain at a distance of $\SI{10}{Mpc}$ from the source in 
the case of the {\tt MIL}, {\tt WhiskyTHC}, and {\tt GRHydro} codes using 
solid purple, dashed orange, and dotted green lines, respectively. 
For a better comparison, we use the same duration signals in all cases, 
corresponding to times ranging from $t\approx\SI{10}{ms}$ to 
$t\approx\SI{20}{ms}$ (with signal duration 
$\tau_{\rm signal}\approx\SI{10}{ms}$), which allows us to focus on the 
post-merger signals.
The vertical 
lines of the same color and pattern mark the peak frequency for each 
case. 
}
\label{fig:GWs_LSBNS}
\end{figure}

\begin{figure*}[htb]
\centering

\begin{tikzpicture}
    \draw (0, 0) node[inner sep=0]{\includegraphics[width=18.3cm]{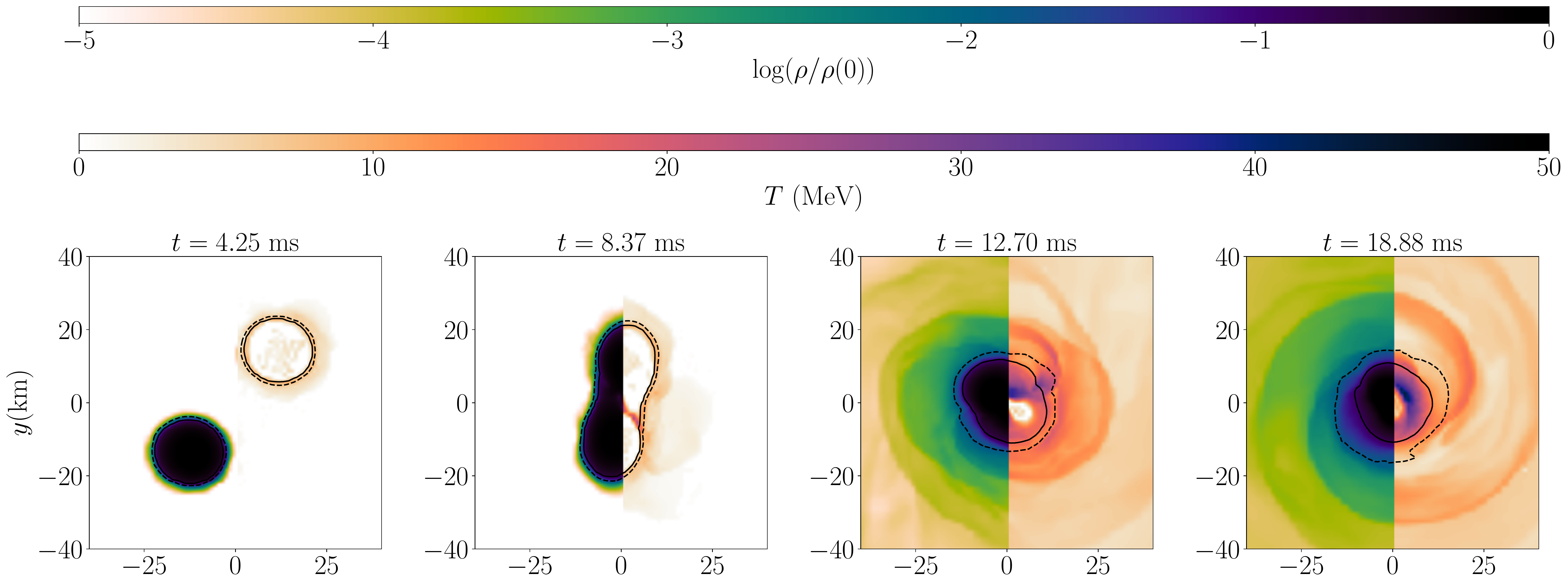}};
    \draw (-7.615, 0.1) node {{\tt MIL}};
\end{tikzpicture}
\begin{tikzpicture}
    \draw (0, 0) node[inner sep=0]{\includegraphics[width=18cm]{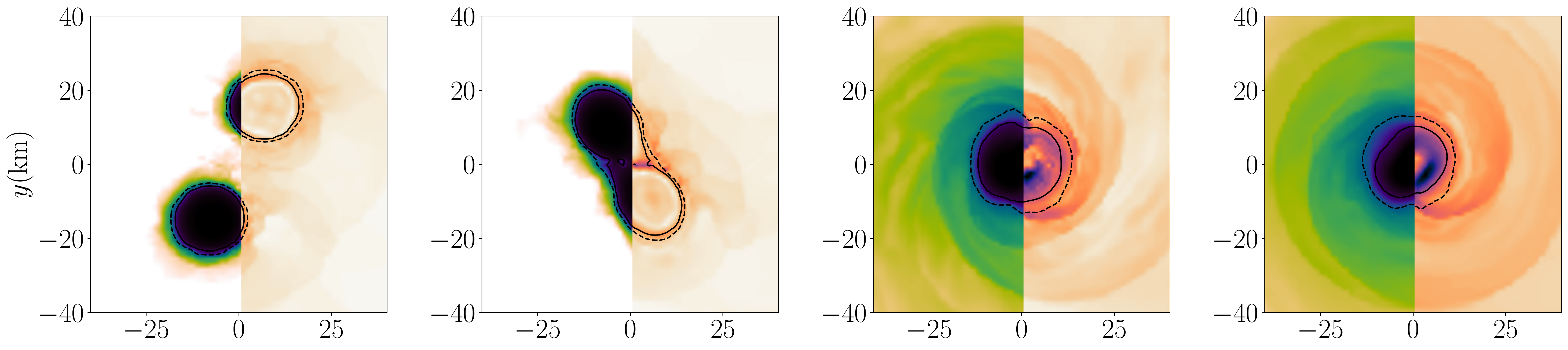}};
    \draw (-7, 1.5) node {{\tt WhiskyTHC}};
\end{tikzpicture}
\begin{tikzpicture}
    \draw (0, 0) node[inner sep=0]{\includegraphics[width=18cm]{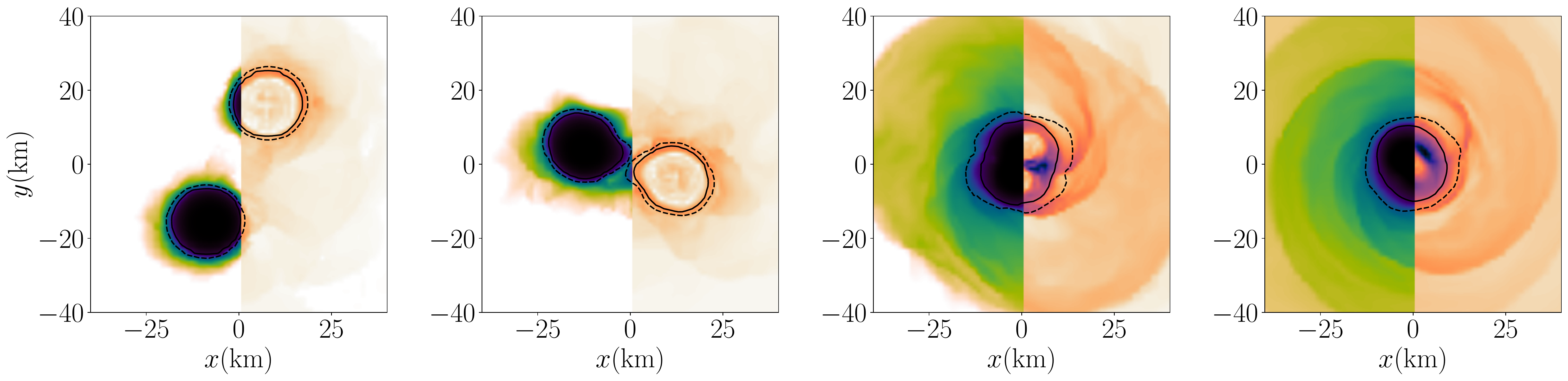}};
    \draw (-7.17, 1.65) node {{\tt GRHydro}};
\end{tikzpicture}
\caption{\textit{Top row:} Equatorial snapshots of the rest mass 
density (left half) and temperature (right 
half) for test LSBNS, in the case of evolution with the {\tt MIL} code. 
From left to right we depict snapshots at times $t=\SI{5.32}{ms}$, 
$t=\SI{8.37}{ms}$, 
$t=\SI{12.70}{ms}$, and $t=\SI{18.88}{ms}$, respectively. The top 
(bottom) colorbar corresponds to the rest mass 
density (temperature). \textit{Middle row:} same as top tow but in 
the case of evolution with the {\tt WhiskyTHC} code. We 
outline regions above which the rest mass density takes on 
values of $\rho_{\rm b}=\SI{e13}{\g\per\cm\cubed}$ and 
$\rho_{\rm b}=\SI{e14}{\g\per\cm\cubed}$ using dashed and 
solid black contours, respectively.
\textit{Bottom row:} same as top row but in the case of evolution 
with the {\tt GRHydro} code.}%
\label{fig:rho_T_inspiral}
\end{figure*}

In this section we consider the evolution of BNS systems built 
using a finite-temperature EOS\@. Specifically, we consider the LS220 EOS
and construct initial data for 
an unmagnetized, equal-mass BNS with ADM mass 
$M_{\rm ADM} = \SI{2.7}{M_\odot}$, and initial separation of 
$\SI{45}{km}$. Our 
evolution grid extends to approximately $\SI{830}{km}$ at the outer 
boundary and consists of 3 additional sets of 6 nested cubes 
(corresponding to 7 levels of refinement); we center one set of cubes at 
the origin of the solution grid, and the other
two sets of cubes are used to track the center of each neutron star 
during the inspiral. 
The grid resolution of each nested cube is half that of the adjacent 
larger cube, 
and the finest-level grid resolution is set such that each NS is resolved 
with at 
least $50$ grid points per radius during the inspiral; generally, the 
grid 
resolution of a given refinement level is 
$dx_l = 2^{(7-l)} R_{\rm NS}/50 $, where 
$l=7(1)$ labels the finest (coarsest) resolution grid and 
$R_{\rm NS}\approx \SI{10.31}{km}$.  We note that 
although these grid resolutions are lower than the standard used for 
production-level runs, such grid resolutions
are routinely considered in the context of BNS mergers as low-resolution 
simulations which are useful in convergence 
studies~\cite{Sekiguchi:2012uc,Hayashi:2021oxy}. We employ these 
grids because we were predominantly
interested in comparisons of different code features. In the context of 
BNS mergers, 
we leave comparisons using high-resolution simulations and a full 
convergence study to future work, since each of the codes we consider has been tested extensively for convergence.
The half-side length of the
finest-resolution 
grid is set to $r_{\rm HSL, 7} = 1.25R_{\rm NS} \approx \SI{12.95}{km}$; 
generally, the half-side length corresponding to a given refinement level  
is 
$r_{{\rm HSL}, l} = 1.25R_{\rm NS} 2^{(7-l)}$. 
In the case of the {\tt MIL} code, we use the HLL Riemann solver.  
In all other cases we use the HLLE Riemann solver. We use PPM for
primitive reconstruction for all evolution codes considered except {\tt WhiskyTHC}, in 
which case we use MP5.

As a way to compare different open-source GRHD codes currently 
available, we consider the evolution of this system with use of the 
{\tt MIL}, {\tt WhiskyTHC}, and {\tt GRHydro} codes.
We present the maximum rest mass density for each evolution in 
Fig.~\ref{fig:rho_LSBNS}, using a purple solid line, 
orange dashed line, and green dotted line for 
the case of {\tt MIL}, {\tt WhiskyTHC}, and {\tt GRHydro}, respectively.  

We identify the merger time $t_{\rm mer}$ as the time where the 
$l=2$, $m=2$ mode of the dominant polarization of the GW strain peaks; 
for reference, we show these GWs in Fig.~\ref{fig:GWs_LSBNS}.
We find merger times of $t_{\rm mer}=\SI{13.57}{ms}$, 
$t_{\rm mer}=\SI{11.72}{ms}$, and $t_{\rm mer}=\SI{12.14}{ms}$
in the case of the
{\tt MIL}, {\tt WhiskyTHC}, and {\tt GRHydro}
evolutions, respectively, which results in an 
approximately $13.6\%$ relative difference, at most, in merger times 
between cases.
In Fig.~\ref{fig:GWs_LSBNS} we show the characteristic 
strain~\cite{Moore:2014lga} extracted
from the simulations in test LSBNS, assuming a source at a distance of 
$\SI{10}{Mpc}$. 
We find 
very similar inspiral GW signals from all codes considered, with relative 
differences in the phase of the gravitational wave strains of at most 
$\sim 2\%$. We find that the largest discrepancy between GW signals 
arises in the immediate post-merger environment. 
The post-merger environment is when the matter is in a highly turbulent and 
dynamical state; in this stage of the merger that the relative difference in 
the phase of the GW strain can reach as high $\sim 20\%$ (when comparing the 
{\tt MIL} and {\tt WhiskyTHC} cases), but is typically closer to 
$\sim 10\%$. After this highly dynamical state, as the remnant begins to 
settle, all codes again produce 
very similar GW signals. The frequency corresponding to peak post-merger 
GW emission $f_{\rm peak}^{2,2}$ is similar between all codes, with the 
{\tt MIL}, {\tt WhiskyTHC}, and {\tt GRHydro} evolutions resulting 
in $f_{\rm peak}^{2,2} = \SI{3.11}{kHz}$, 
$f_{\rm peak}^{2,2}=\SI{3.06}{kHz}$, and 
$f_{\rm peak}^{2,2}=\SI{3.13}{kHz}$.

In the top, middle, and lower panels of Fig.~\ref{fig:rho_T_inspiral}
we show snapshots of the rest mass density and temperature in the
equatorial plane at key points during the merger for the case of {\tt
  MIL}, {\tt WhiskyTHC}, and {\tt GRHydro}, respectively.  For each
evolution code considered we depict 4 snapshots, ordered by time from
left to right at different stages of the merger corresponding to times
$t\in(4.25,8.37,12.70,18.88) \text{ ms}$.  In the left-most panels in
Fig.~\ref{fig:rho_T_inspiral} we depict the system at a time during
the inspiral.  We generally find that the binary components remain
cold during the inspiral, with typical temperatures inside the stars
of $T\lesssim 1-\SI{2}{MeV}$.  Temperatures of
$\mathcal{O}(\SI{1}{MeV})$ within the stars are expected during the
inspiral~\cite{raithel2021realistic}, and may be due to the growth
of numerical error seeded at the level of initial data or interpolation 
errors from the EOS tables.  We find that
the stellar atmospheres become significantly warm as the stars
inspiral, reaching temperatures of $T_{\rm atmo}\approx
5-\SI{10}{MeV}$ before even the first orbit is complete (note that we employ a density cutoff to define the 
atmosphere, such that all matter with below 
$\rho_{\rm b, atmo}\approx \SI{6.17e+07}{\g\per\cm\cubed} \approx 10^{-7}
\rho_{\rm b, max}(0)$ is set to the atmosphere as detailed in 
Sec.~\ref{subsec:CH7_C2P}). This temperature profile (cold
temperatures of $T\approx \SI{1}{MeV}$ within the stars and higher
temperatures $T_{\rm atmo} \approx \SI{10}{MeV}$ in the atmospheres)
is maintained during the entire inspiral up to merger, as depicted by
the leftmost panels in Fig.~\ref{fig:rho_T_inspiral}. We find that,
during the inspiral, the warmest temperatures develop in the case of
the {\tt GRHydro} and {\tt WhiskyTHC} evolutions, reaching as high as
$T\approx \SI{5}{MeV}$ in the bulk of the star and $T_{\rm atmo}
\approx \SI{15}{MeV}$ in the atmosphere. These cases also exhibit
warm, low-density clouds which develop around the binary components
throughout the inspiral. However, the
atmosphere is dynamically unimportant.

\begin{figure*}[htb]
\centering
\includegraphics{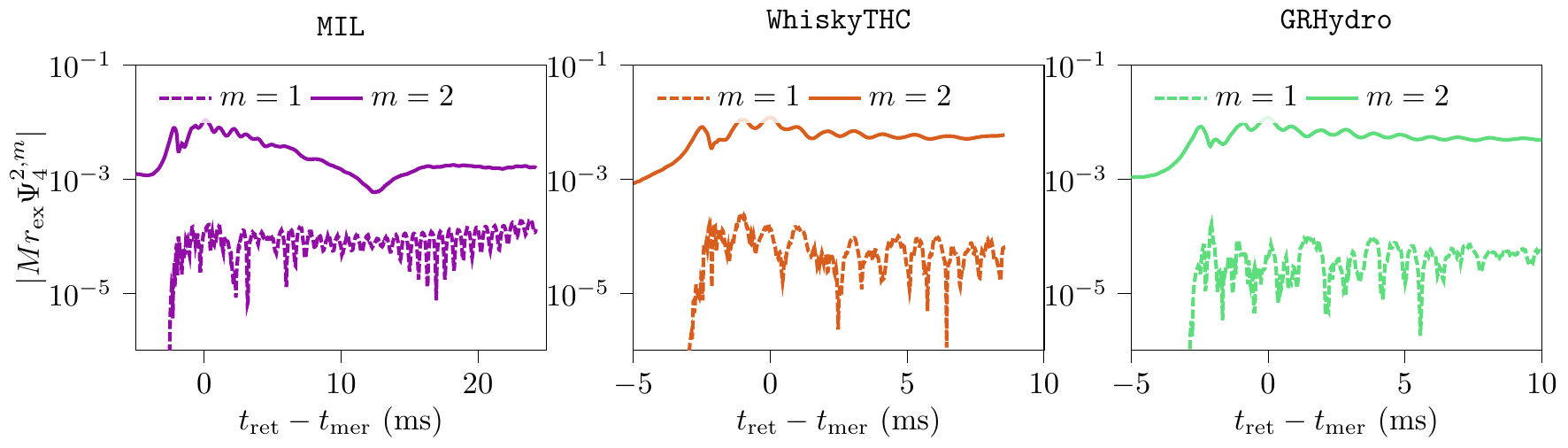}
\caption{Dominant coefficients of 
the spin-weighted decomposition of the Newman-Penrose scalar $\Psi_4^{2,m}$ 
(scaled with the extraction radius
$r_{\rm ex}$ and ADM mass $M$) as a function of retarded time 
$t_{\rm ret}$ (shifted by the time of merger $t_{\rm mer}$) for the $l=2$,
$m=1$ (dashed lines) and $l=2$, $m=2$ (solid lines) GW modes for test 
LSBNS\@. In the left, center, and right panels we show results for evolution
with {\tt MIL} (purple lines), {\tt WhiskyTHC} (orange lines), and {\tt 
GRHydro} (green lines), respectively}%
\label{fig:Psi4_22_21}
\end{figure*}

In the second-from-left column in Fig.~\ref{fig:rho_T_inspiral} we depict 
the systems at a time near merger. At this point in the 
evolution the first significant shock heating 
happens as the cores of the two stars collide. In all cases, temperatures 
in the shear layer between the two stellar cores can climb to 
$T \approx 30-\SI{40}{MeV}$. As the two cores 
continue to orbit and merge, the warm shocked material is redistributed 
toward the outside of the merger remnant, 
resulting in a warm envelope of $T \approx\SI{10}{MeV}$ surrounding the 
merged core which can reach temperatures of $T\approx \SI{40}{MeV}$. We 
depict such states in the third column 
of Fig.~\ref{fig:rho_T_inspiral}, 
which roughly correspond to a few milliseconds after merger. 
Finally, in the 
rightmost panels of Fig.~\ref{fig:rho_T_inspiral} we show the state of 
the merger remnants at a time corresponding to approximately 
$7-\SI{9}{ms}$ 
after merger. As the merger remnant settles, the final configurations 
approach a warm central object with temperatures $T\approx \SI{10}{MeV}$. 
We find that in all cases this central object is surrounded by a 
distribution of matter which resembles a ring 
of hot material (with temperatures ranging $T\approx 40-\SI{50}{MeV}$),  
which becomes cooler with increased radial distance from the central 
configuration (dropping to temperatures $T\approx 10-\SI{20}{MeV}$). 

\begin{figure}[htb]
\centering
\includegraphics{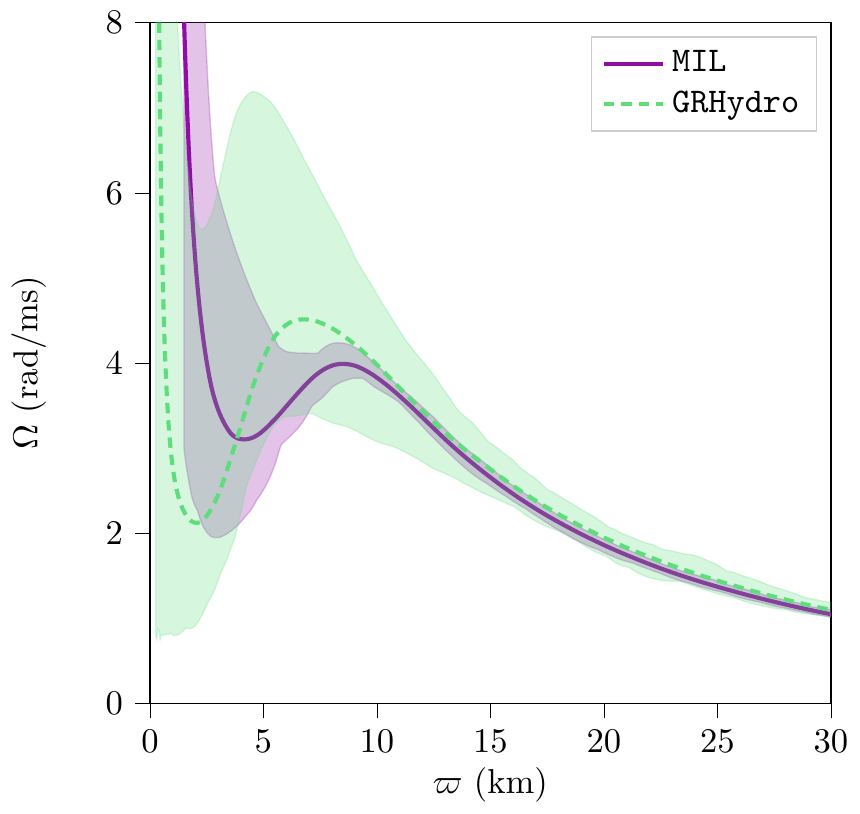}
\caption{Time-averaged, azimuthally-averaged angular velocity as a 
function of radial distance on the equatorial plane for 
the case where we use the {\tt MIL} (purple solid line) and 
{\tt GRHydro} (dashed green line) codes in test LSBNS\@. We
time-average over a period of approximately $\SI{10}{ms}$, ranging from 
simulation times of $t\approx \SI{20}{ms}$ to $t\approx \SI{30}{ms}$. The 
purple and green contours depicts the range of values for the 
instantaneous angular velocity in the time window we consider for the 
{\tt MIL} and {\tt GRHydro} evolutions, respectively. The profiles 
indicate that the remnant is differentially rotating by the end of 
simulation. }%
\label{fig:angvel}
\end{figure}

For a clear view of remnant properties, we consider rest mass density 
cutoffs which 
allow us to identify regions of the remnant. We 
identify the remnant itself and its core as regions in the post-merger 
environment above 
$\rho_{\rm b} \geq \SI{e13}{\g\per\cm\cubed}$ and 
$\rho_{\rm b} \geq \SI{e14}{\g\per\cm\cubed}$, 
respectively~\cite{Kiuchi:2017zzg}. We highlight the remnant and its core 
using dashed and solid black lines in Fig.~\ref{fig:rho_T_inspiral}, 
respectively.
We find that, in all cases, immediately following the merger a dense core 
is formed near the origin which typically 
extends to coordinate distance of $\sim\SI{15}{km}$ from the origin and is surrounded by lower 
density material which extends to coordinate distance $\sim \SI{20}{km}$. This HMNS remnant 
is surrounded by a low-density, disk-like structure.

The two key factors that determine the final fate of a BNS merger remnant 
are the remnant rest mass and the maximum rest mass of a uniformly
rotating NS (i.e., the `supramassive' limit mass) $M_{\rm supra}$ for the
given EOS\@. We note that, for hadronic EOSs like the ones considered in
this work, the value of $M_{\rm supra}$ is closely
related to the value of the maximum rest mass of a non-rotating star
$M_{\rm TOV}$ and is expected to be about
$20\%$ larger
$M_{\rm supra} \approx 1.2 M_{\rm TOV}$~\cite{CST94a, CST94b,
Lasota1996,Morrison2004,Breu2016}  (although generally this mass increase
can reach up to about $35\%$~\cite{Bozzola:2019tit}); in the case of the
LS220 EOS, $M_{\rm supra}=\SI{2.83}{M_\odot}$, and the total system rest
mass in test LSBNS is $M_{\rm b}=\SI{2.89}{M_\odot}$. One of the main sources of support
against gravitational collapse in BNS merger remnants comes from the
large amounts of differential rotation in the
system (thermal support can also contribute significantly against
gravitational collapse~\cite{raithel2021realistic}). Differential rotation can
greatly increase the maximum mass a star can support~\cite{Ansorg:2009, Bozzola:2017qbu, Espino2019},
but, as differential rotation is removed from the
system (for example, 
via braking by magnetic fields~\cite{Shapiro_2000, Shibata:2021bbj}), the
remnant will approach a uniformly rotating NS\@. If the total remnant mass
is below $M_{\rm supra}$ for a given EOS, we may expect a long-lived NS 
remnant~\cite{Margalit2017}. 
On the other hand, for systems with $M_{\rm tot} > M_{\rm supra}$, we may 
expect collapse. In all cases, we observe a transient HMNS remnant, as 
expected for this total binary mass and EOS\@.
A key difference in the post-merger evolution for the simulations we 
consider is in the survival time of the HMNS remnant. 
In the {\tt MIL} case, we find 
that the remnant does not collapse to a black hole (BH) by the end of 
evolution ($t_{\rm fin} \sim 40 \text{ ms}$). 
On the contrary, the remnant collapses
after $t_{\rm coll} \approx \SI{21.6}{ms}$ and $t_{\rm coll} \approx 
\SI{25.2}{ms}$ in the case of the {\tt WhiskyTHC} and {\tt GRHydro} 
evolutions, respectively (demonstrated by the sharp rise in the rest mass 
density in Fig.~\ref{fig:rho_LSBNS} and corroborated by the collapse of the 
lapse function $\alpha$ in both cases).

The remnant survival time $\tau_{\rm remnant}$ is an important quantity 
which significantly affects several observables associated with BNS 
mergers. For instance, neutrino irradiation from a metastable HMNS may 
affect ejecta properties, which would in turn affect the KN 
signal~\cite{Metzger:2014ila, Margalit2017, Holmbeck:2020ueu}. 
Additionally, $\tau_{\rm remnant}$ is expected to affect the 
nucleosynthesis in accretion 
disk outflows, with shorter HMNS lifetimes resulting in relatively low 
lanthanide and actinide abundances~\cite{Lippuner:2017bfm} due to the 
shorter neutrino irradiation times~\cite{Metzger:2014ila, Martin:2014ybv, 
Martin:2015hxa,Miller:2019mfl}.
The significant differences in remnant survival times 
$\tau_{\rm remnant}$ exhibited by each code suggests that additional 
sutdies must be carried out which rely on the value of $\tau_{\rm remant}$
extracted from NR simulations. The remnant survival time may sensitively 
depend on the EOS~\cite{Hotokezaka:2013iia, Lucca:2019ohp}, 
neutrino treatment (relevant for cooling and emergent viscous 
effects)~\cite{Duez:2004nf, PhysRevD.95.123003, Palenzuela:2022kqk}, 
magnetic field treatment (relevant for angular momentum 
transport)~\cite{Shapiro_2000, Radice:2017zta, Shibata:2021bbj, 
Palenzuela:2022kqk}, and (as 
suggested by the tests presented here) differences in the implementation of
algorithmically equivalent numerical methods. Additional, in-depth
studies remain to be carried to systematically understand all of the
aforementioned effects on $\tau_{\rm remnant}$.

 The survival time of the remnant can also play a role in
the development of different fluid instabilities. Fluid instabilities
that uniquely develop in the post-merger environment - such as the
one-arm spiral instability and magneto-rotational instability (MRI) -
would be arrested upon gravitational collapse to a black hole. For
instance, a potential observable signature which arises from the
one-arm spiral instability~\cite{Centrella:2001xp,Saijo2003} is the
long-lived, sustained powering of the $l=2$, $m=1$ GW
mode~\cite{Paschalidis:2015mla}, which typically has half the
characteristic frequency of the initially dominant but decaying $l=2$,
$m=2$ GW mode~\cite{Paschalidis:2015mla,Radice:2016gym}.  Over
relatively long timescales, and as long as the post-merger massive
neutron star remnant has not collapsed to a black hole, the one-arm
spiral mode can continuously power the emission of the $l=2$, $m=1$ GW
mode~\cite{Paschalidis:2015mla}; simulations which produce shorter
remant survival times may not allow the one-arm spiral instability to
develop. In Fig.~\ref{fig:Psi4_22_21} we show the $l=2$, $m=2$ (using
solid lines) and $l=2$, $m=1$ (using dashed lines) GW modes for test
LSBNS\@.  In all simulations we see increased emission of the $m=1$
mode at a time close to merger, suggesting the potential early
development of the one-arm spiral instability. In the {\tt WhiskyTHC}
and {\tt GRHydro} evolutions, the amplitude of the $m=1$ mode remains
roughly constant until collapse, at which point the emission in this
mode ceases. However, in the case of the {\tt MIL} evolution, which
has not collapsed by the end of simulation, GW emission in the $m=1$
mode continues and is further amplified, while emission in the $m=2$
mode decays.  Due to the different remnant collapse times, the $m=1$
mode is allowed to continue growing in the {\tt MIL} evolution, but
not in the {\tt WhiskyTHC} and {\tt GRHydro} evolutions.

Another set of instabilities that is relelvant in the post-merger
environment, and are thereby affected by the remnant collapse time, are
magnetic instabilities\@. In particular, the MRI can develop in the
strongly differentially rotating environments following a BNS
merger~\cite{Siegel:2013nrw, PhysRevD.90.041502, Kaplan2014,
  Guilet2016, Ruiz_2016, Paschalidis2017}.  As long as differential
rotation with an appropriate profile is sustained in a massive neutron
star post-merger remnant, the MRI may develop and produce exponential
magnetic field amplification
(see~\cite{PaschalidisStergioulas2017,Ciolfi:2020cpf} and references
therein).  To consider the amount of differential rotation in the
post-merger remnants depicted in Fig.~\ref{fig:rho_T_inspiral} we
calculate the approximate angular velocity in the orbital plane
$\Omega = v^\phi$, using Eq.~\eqref{eq:angvel}.  In
Fig.~\ref{fig:angvel} we show the radial profile of the angular
velocity $\Omega(\varpi)$, averaged along the azimuthal direction and
in time, for test LSBNS in the case of evolution with the {\tt MIL}
code (depicted using purple lines and contours) and the {\tt GRHydro}
code (depicted using green lines and contours). We time-average over a
time window of approximately $\SI{10}{ms}$ ranging from
$t\approx\SI{20}{ms}$ to $t\approx\SI{30}{ms}$ for the {\tt MIL} case
and from $t\approx\SI{14}{ms}$ to $t\approx\SI{24}{ms}$ in the {\tt
  GRHydro case}.  We find that by the end of each simulation the
remnants still exhibits significant amounts of differential rotation.
The supramassive limit for EOS LS220 falls below the total system
mass. Therefore, we may reasonably expect a delayed collapse once the
configuration is no longer supported by differential rotation. The
timescale on which differential rotation is removed in BNS merger
remnants is typically $\sim
\mathcal{O}(\SI{100}{ms})$~\cite{Ciolfi:2019fie,Ciolfi:2020wfx}, which
is significantly longer than the timescales probed by our dynamical
simulations. However, such arguments assume that typically a large
fraction of the angular momentum must be lost or transport for
collapse to proceed, which is not generally the case. Therefore,
although we may expect the same final remnants -- a black hole -- in
all of our simulations for test LSBNS, the timescales over which this
happens varies significantly across codes. We note that similar
amounts of differential rotation are observed in the {\tt GRHydro}
case
as well, suggesting that it was not the lack of differential rotation
support which led to the collapse of the post-merger remnants in that
case.  As the massive NS remnants for both the {\tt MIL} and {\tt
  GRHydro} cases exhibit a significant amount of differential rotation
by the end of simulation (or until collapse in the {\tt GRHydro}
case), we may expect the MRI to develop, if sufficiently resolved, in
analogous simulations which include magnetic fields. However, our
results suggest that the MRI could have a significantly longer time to
develop in evolutions with the {\tt MIL} code, due to a potentially
significantly longer remnant lifetime than we can probe. On the other
hand, the magnetic field can also reduce or prolong the lifetime of
the remnant. The comparison cases presented in this section suggest
that understanding the error introduced by using different numerical
methods in NR simulations of BNS mergers is crucially important to the
questions of magnetic field amplification mechanisms, fluid
instabilities, and observables associated with such systems.

\section{Conclusion and Outlook}\label{sec:CH7_conclusion}
In this work we have demonstrated the capabilities of the \texttt{MIL} 
code, a code which builds upon the current open source version of 
\texttt{IllinoisGRMHD} by including necessary changes for 
microphysical EOS compatibility. We outlined the algorithmic 
structure of  \texttt{IllinoisGRMHD} and expanded on the additions 
that \texttt{MIL} provides. Specifically, we 
discussed the implementation of the electron fraction advection equation 
and the implementation of new 
state-of-the-art conservatives-to-primitives solvers within the code. The 
main products of these aforementioned additions are the 
\texttt{MIL} code itself and a new standalone thorn from primitives 
recovery, \texttt{ConservativeToPrimitive},  
which is in principle able to interface with any GRMHD code that uses 
the \texttt{Cactus} infrastructure.
Over a barrage of tests we established that the \texttt{MIL} code works 
as well as other publicly available GRMHD codes with similar 
capabilities. 
Crucially, for the systems considered, the \texttt{MIL} code  agrees to a 
large extent with the \texttt{Spritz}
code {\tt GRHydro}, and {\tt WhiskyTHC} codes.

Throughout the work we considered the numerical solutions provided 
by \texttt{MIL} for a wide range of 
astrophysical relevance. In particular, we consider several TOV and BNS
configurations, all of which provide stringent tests 
of the \texttt{MIL} code. Where relevant, we compare with 
other open source GRMHD codes. We find general agreement and similar 
convergence properties to the 
\texttt{GRHydro}, \texttt{Spritz}, and {\tt WhiskyTHC} codes, as well as 
to the current open source version of {\tt IllinoisGRMHD} 
(referred to as \texttt{OIL} throughout the work). In some cases we find 
improved performance for \texttt{MIL} when compared 
to \texttt{GRHydro} (a property which is carried over from the 
\texttt{OIL} code as explored in~\cite{Etienne:2015cea}). We find 
general quantitative agreement between all codes considered in scenarios 
where the matter remains at low temperatures throughout the 
evolution. The largest differences observed between the results of each 
code appear during highly dynamical scenarios such as during BNS mergers 
and in the immediate post-merger evolution. Although we find agreement in 
the post-merger peak frequency of the characteristic strain, other 
quantitative differences 
arise in these highly dynamical scenarios, likely due to the differences 
in the implementation of conservative-to-primitive algorithms. 
For example, some differences we highlight include: (1) the {\tt GRHydro} 
and {\tt WhiskyTHC} codes leading to higher temperatures than the {\tt MIL}
code in a BNS merger scenario, typically by up to 
$\Delta T \approx \SI{10}{MeV}$; (2) merger times which differ by 
approximately $10 \%$ between codes; (3) significantly different survival 
times of the post-merger remnant; (4) large oscillations in the rest mass 
density for {\tt GRHydro} when comparing to other codes, and general lack 
of quantitative agreement in the post-merger evolution; (5) 
better preservation of the initial magnetic field structure during the 
inspiral in the {\tt OIL} and {\tt MIL} codes when compared to the {\tt 
Spritz} code.

Given that all codes considered allow some freedom in the flux treatment 
and reconstruction scheme, we ensure in all our tests that these two 
aspects of the codes have maximal algorithmic overlap. However, 
even if two codes employ the same algorithm, the implementation 
methods will be different in each case. These differences in 
algorithmic implementation may lead to the quantitative differences noted 
throughout the work. Despite the quantitative differences highlighted, we 
emphasize that all codes nevertheless produce qualitatively consistent 
(magneto)hydrodynamics evolutions 
and result in quantitatively similar gravitational waves. 
We note that 
the \texttt{MIL} code provides additional 
freedom over~\texttt{GRHydro}, \texttt{Spritz}, and {\tt WhiskyTHC} in 
the primitives inversion algorithm used during any given 
simulation through the thorn \texttt{ConservativeToPrimitive}, allowing 
for suitable comparisons with different codes and the ability to use the 
optimal conservative-to-primitive solver for the problem at hand. 
Additionally, the use of several solvers 
within~\texttt{ConservativeToPrimitive} for a given simulation allows for 
a robust and efficient primitives recovery solution in the case of finite 
temperature EOSs, regardless of the variable space probed. 

The updates included in \texttt{MIL} make possible the investigation of 
important, open questions surrounding BNS mergers. In particular, 
\texttt{MIL} will allow us to investigate the interplay between the 
effects of magnetic fields and microphysical and finite temperature EOSs 
in BNS mergers, the calculation of nucleosynthesis yields associated with
BNS mergers, and provides the groundwork for the inclusion of neutrino 
effects within \texttt{IllinoisGRMHD}. Future work utilizing the 
full capabilities of \texttt{MIL} will investigate all of the 
aforementioned phenomena. For instance, the tests 
presented in Sec.~\ref{subsec:long_term} will be expanded upon for a 
full investigation of the interplay between 
magnetic field and realistic EOS effects. Additionally, because the 
\texttt{MIL} code provides a dynamical 
description of several fluid and thermodynamic quantities -- including 
the rest mass density, temperature, 
specific entropy, and electron fraction -- its output may be used with 
nuclear reaction network codes such as 
\texttt{SkyNet}~\cite{Lippuner_2017} which only require such quantities 
as input. In turn, nuclear reaction 
networks provide insight into the nucleosynthesis rates associated with 
different stages of BNS mergers. Full 
GRMHD simulations of BNS mergers with magnetic field constrained 
transport have not received much attention in the 
context of nucleosynthesis, and this is now possible with \texttt{MIL}. 
Finally, because the electron fraction is known at all 
times during evolutions with \texttt{MIL}, the code may be expanded 
further to include neutrino effects. For 
instance, one may include neutrino leakage by interfacing \texttt{MIL} 
with the \texttt{ZelmaniLeak} thorn~\cite{Ott_2009}, or include
more accurate descriptions of neutrino transport that move 
beyond leakage~\cite{Foucart_2020,Vsevolod:2020pak,
Nedora:2020qtd,Weih_2020,Radice:2021jtw}. Such an 
extension to \texttt{MIL} will help elucidate the combined effect of 
magnetic and neutrino driven outflows after a BNS merger. 
Specifically, neutrino effects are expected to have a significant role on 
the electron fraction of the outflow~\cite{Radice:2021jtw}. 
The constrained transport magnetic field evolution and finite temperature 
description of \texttt{MIL} coupled with accurate neutrino transport 
could provide some of the most detailed insight into the physics 
following a BNS merger. We note that another recent 
extension to {\tt IllinoisGRMHD} 
similar to ours (albeit using alternative methods and implementations) 
was considered in~\cite{Werneck:2022exo}. A detailed comparison of the 
codes considered in this work along with the results presented 
in~\cite{Werneck:2022exo} is a crucial step toward 
understanding systematic errors in NR simulations of BNS mergers.
We hope to consider these and other investigations in future work 
with the {\tt MIL} code. We conclude the present work with a note about 
the availability of the {\tt MIL} code. The source code for {\tt MIL}, 
along with the supporting thorns, is publicly available at 
{\tt https://github.com/pilambdaepsilon/UAThorns}.We also provide examples 
of parameter files, initial data, and equation of state files for all of 
the examples considered in this work. 

\section{Acknowledgments}
This work was supported by NSF Grants PHY-1912619 and PHY-2145421 to the
University of Arizona. PE also acknowledges support from NSF Grant PHY-2020275
(Network for Neutrinos, Nuclear Astrophysics, and Symmetries (N3AS)). During
completion of this work, PE was in part supported by the Marshall Foundation
Dissertation Scholarship. GB is supported by NASA Grant 80NSSC20K1542 to the
University of Arizona. We wish to thank Bruno Giacomazzo, Federico 
Cipolletta, and David Radice for discussions.
Simulations were in part performed at the {\tt Comet} and {\tt Expanse} clusters
at SDSC, and the {\tt Stampede2} cluster at TACC through XSEDE grant
TG-PHY190020. Simulations were also performed on the \texttt{ElGato},
\texttt{Ocelote}, and {\tt Puma} clusters at the University of Arizona.
\texttt{kuibit}~\cite{kuibit} is based on \texttt{NumPy}~\cite{NumPy},
\texttt{SciPy}~\cite{SciPy}, and \texttt{h5py}~\cite{h5py}.

\bibliography{ref}

\end{document}